\documentclass[aps,amsmath,amssymb,nofootinbib,longbibliography]{revtex4-2}

\usepackage[letterpaper, margin=0.75in]{geometry}
\usepackage{amsmath}
\usepackage{amssymb}
\usepackage{amsfonts}
\usepackage{xcolor}

\usepackage{graphicx}
\usepackage{subcaption}

\definecolor{lcolor}{rgb}{0.5,0,0}
\definecolor{citcolor}{rgb}{0,0.3,0.0}
\usepackage[breaklinks,colorlinks,urlcolor=blue,citecolor=citcolor,linkcolor=lcolor]{hyperref}

\graphicspath{ {.} }

\newcommand{\ket}[1]{\left\vert #1 \right\rangle}
\newcommand{\bra}[1]{\left\langle #1 \right\vert}

\newcommand{\exv}[1]{\left\langle #1 \right\rangle}
\newcommand{\tr}{\text{tr}\,}

\newcommand{\intddkg}{\int \frac{\mathrm{d}^2k_g}{16\pi^3}}
\newcommand{\nc}{N_C}
\newcommand{\vn}{{\rm vN}}
\newcommand{\dd}{{\rm d}}

\begin{document}

\title{Quantum entanglement correlations in double quark PDFs}

\author{Adrian Dumitru}
\email{adrian.dumitru@baruch.cuny.edu}
\affiliation{Department of Natural Sciences, Baruch College, CUNY,
17 Lexington Avenue, New York, NY 10010, USA}
\affiliation{The Graduate School and University Center, The City University
  of New York, 365 Fifth Avenue, New York, NY 10016, USA}

\author{Eric Kolbusz}
\email{ekolbusz@gradcenter.cuny.edu}
\affiliation{Department of Natural Sciences, Baruch College, CUNY,
17 Lexington Avenue, New York, NY 10010, USA}
\affiliation{The Graduate School and University Center, The City University
  of New York, 365 Fifth Avenue, New York, NY 10016, USA}

\begin{abstract}
  Methods from Quantum Information Theory are used to scrutinize
  quantum correlations encoded in the two-quark density matrix over
  light-cone momentum fractions $x_1$ and $x_2$. A non-perturbative
  three quark model light-cone wavefunction predicts
  significant non-classical correlations associated with the ``entanglement
  negativity'' measure for asymmetric and small quark momentum
  fractions.  We perform one step of QCD scale evolution of the entire
  density matrix, not just its diagonal (dPDF), by computing
  collinearly divergent corrections due to the emission of a gluon.
  Finally, we present first qualitative numerical results for
  single-step scale evolution of quantum entanglement correlations in
  double quark PDFs. At a higher $Q^2$ scale, the non-classical correlations
  manifest in the dPDF for nearly symmetric momentum fractions.
\end{abstract}

\maketitle
\tableofcontents

\section{Introduction}
Multi parton interactions (MPI) may occur in high energy proton-proton
or proton-nucleus collisions whereby two or more partons from the
projectile proton undergo a hard scattering off the target. The
simplest such MPI process corresponds to double parton scattering
which involves double parton distributions
(dPDFs)~\cite{Korotkikh:2004bz,Gaunt:2009re,Blok:2010ge,Diehl:2011yj};
we refer to ref.~\cite{Diehl:2017wew} for a review of the theory of
double parton scattering. A setup for the computation of double parton
distributions via lattice QCD has been described in
ref.~\cite{Jaarsma:2023woo}.  dPDFs describe the joint distribution of
two partons with light-cone (LC) momentum fractions $x_1$ and $x_2$,
respectively, in the proton.

dPDFs provide very interesting insight into parton {\em correlations}
in the proton.  Indeed, a number of authors have concluded that a
simple factorized dPDF, lacking correlations when $x_1+x_2<1$, of the
form
\begin{equation} \label{eq:factorized-dPDF}
f_{ij}(x_1, x_2, Q^2) = f_i(x_1,Q^2) \, f_j(x_2,Q^2) \,
\Theta(1-x_1-x_2)
\end{equation}
does not provide a good approximation to the full
dPDF~\cite{Korotkikh:2004bz,Gaunt:2009re,Blok:2010ge,Broniowski:2013xba,Golec-Biernat:2014bva}.
Here, $i$ and $j$ denote parton flavors, and it has been assumed that
the MPI involve a single hard scale $Q^2$, for simplicity.  We shall
first consider the double quark PDF at a low virtuality $Q_0^2$ of
order a hadronic scale; henceforth $Q_0^2$ will be omitted from the
arguments of the dPDFs.  If the dPDF at the initial scale $Q_0^2$
lacks correlations, then factorization of the dPDF is preserved by QCD
evolution.  In sec.~\ref{sec:rho_evol} we perform a single evolution
step towards higher $Q^2$ of the entire density matrix including its
off-diagonal elements\footnote{We note in passing that
ref.~\cite{Armesto:2019mna} considered small-$x$ evolution, in the
high gluon density limit, of the complete density matrix for soft
gluons.}. This provides first qualitative insight into double parton
quantum correlations encoded in the density matrix at $Q^2>Q_0^2$.  \\

A variety of initial conditions for dPDFs with correlations have been
proposed. The model by Gaunt and Sterling (GS)~\cite{Gaunt:2009re} for
the valence quark dPDFs corresponds to
\begin{equation} \label{eq:GS-dPDF}
f_{ij}(x_1, x_2) = f_i(x_1) \, f_j(x_2)\,
\frac{(1-x_1-x_2)^2}{(1-x_1)^{2+n}\, (1-x_2)^{(2+n)}}
\end{equation}
with $n=0.5$. It is understood that the support of this function is
restricted to $x_1, x_2 > 0$ and $x_1+x_2 \le 1$.  In this model
correlations arise mainly due to the phase space restriction and the
dPDF approaches a factorized form as $x_1+ x_2 \ll 1$.

Another model (``model II'' in~\cite{Broniowski:2013xba}) for correlated 
valence quark dPDFs has been proposed by Broniowski and Arriola (BA)
\begin{equation} \label{eq:BA-dPDF}
f_{ij}(x_1, x_2) = f_i(x_1) \, f_j(x_2)\,\frac{725}{28}
\frac{(x_1+x_2)^2}{(1-x_1)(1 + 6x_1 + 16x_1^2 + 6x_1^3 + x_1^4) (1-x_2)(1 + 6x_2 + 16x_2^2 + 6x_2^3 + x_2^4)}\ .
\end{equation}
This model respects the quark number and momentum sum rules.
However, here $f_{ij}(x_1, x_2) / f_i(x_1) \, f_j(x_2)$ does not vanish
as $x_1+x_2 \to 1$.

In passing we mention also ref.~\cite{Golec-Biernat:2015aza} who
proposed a construction of initial conditions (at a low $Q^2$ scale)
for double gluon PDFs from single gluon PDFs. Here we focus on quark dPDFs.

Valence double quark PDFs have also been computed in a ``bag model''
of the proton~\cite{Chang:2012nw}. In a ``rigid bag'' approximation
the dPDF factorizes and there are no correlations. However, other
treatments of the bag do break factorization into single quark
PDFs. In the bag model we do not have a simple analytic form of the
dPDF like in the two models mentioned above, and so we will not
consider it further in this paper.  We do note, however, that this
model in principle does provide the complete two-quark density matrix
including its off-diagonal elements. Hence, it may be possible to
distinguish quantum vs.\ classical correlations using methods similar
to those employed here.

Rinaldi {\it et al.} have derived double quark distributions from a
constituent quark model light-cone wavefunction (LCwf) of the
proton~\cite{Rinaldi:2014ddl,Rinaldi:2016mlk,Rinaldi:2016jvu} by
``tracing over'' the remaining quark. This way, all sum rules are
satisfied by construction. A rather similar approach is presented
here. However, our formulation also provides the {\em off-diagonal}
elements of the two-quark density matrix. While not required for the
construction of the dPDF, knowledge of the entire density matrix is
needed to determine the nature of correlations, such as quantum
vs.\ classical, by the application of methods from Quantum Information
Theory (QIT).

While the present focus is on double quark PDFs as a function of LC
momentum fraction, we note that ``flavor interference dPDFs'' have
recently been mentioned in the
literature~\cite{Reitinger:2024ulw}. These correspond to off-diagonal
elements of the two-quark density matrix over flavor, analogous to the
off-diagonal elements of the two-quark density matrix over $x$
considered here. As already mentioned above, the novel aspect here is
to show how specific methods from QIT can be employed to scrutinize
the classical vs.\ quantum nature of double quark correlations.

There has been great interest recently in understanding various
aspects of entanglement of quarks and gluons in the proton, initiated
primarily by the work of Kharzeev {\it et
  al.}~\cite{Kharzeev:2017qzs,Kharzeev:2021nzh,Kharzeev:2021yyf} as
well as Kovner {\it et
  al.}~\cite{Kovner:2015hga,Kovner:2018rbf,Armesto:2019mna,Duan:2021clk,Dumitru:2023qee}. Ref.~\cite{Hentschinski:2024gaa},
which focuses on the rapidity evolution of the entanglement entropy,
provides a list of additional references on this topic.  Moreover,
Hatta {\it et al.} have recently analyzed entanglement of spin and
orbital angular momentum degrees of
freedom~\cite{Hatta:2024lbw,Bhattacharya:2024sno}.

The goal of this paper is to explore the application of
QIT methods to reveal the potential presence of quantum correlations
in double quark PDFs.  In the following sec.~\ref{sec:Detection_qC}
we identify a measure for quantum correlations and a method for
suppressing them in order to study the effect on the dPDF. In
sec.~\ref{sec:dPDF_qqq} we introduce a simple, non-perturbative
constituent quark model for the effective light-cone wave function
of the proton which provides us with the entire two-quark density
matrix including off-diagonal elements. Furthermore, we identify a
valid ``partitioning'' of the reduced density matrix, in terms of
unconstrained variables for the quark momentum fractions, in order
to be able to study subsystem correlations with QIT methods.  We
discuss the convergence of the eigenspace of the discretized,
partially transposed density matrix towards an infinite dimensional
null space times a finite dimensional subspace for non-zero
eigenvalues.  Sec.~\ref{sec:rho_evol} is devoted to the
implementation of the first step of perturbative QCD scale evolution
of the {\em entire} two-quark density matrix. Only the diagonal
elements evolve in terms of convolutions of splitting functions with
dPDFs but such treatment is insufficient for an analysis of the
scale evolution of quantum correlations.

\subsection{Detection of Quantum Correlations}
\label{sec:Detection_qC}

Our goal is to identify specific correlations in the two-quark density
matrix (and hence, the dPDF) which cannot arise classically and
therefore must be quantum in nature. General information theoretical
analysis of correlations is basis-independent and so requires the eigenvalues of the
density matrix, meaning that knowledge of the entire matrix, not just its diagonal, is necessary. We will
briefly summarize the relevant QIT theory for the sake of completeness.

Given two systems (e.g. quark 1 and quark 2) separately described by 
density matrices $\rho^{A}$ and $\rho^{B}$ in Hilbert spaces
$\mathcal{H}^{A}$ and $\mathcal{H}^{B}$, there are multiple
possibilities for the behavior of the combined system. If the
subsystems are disjoint, then the total density matrix $\rho =
\rho^{A} \otimes \rho^{B}$ is called a product state and clearly has
no correlations between the two systems. In the case of two quarks,
this corresponds to a factorized dPDF as in
\eqref{eq:factorized-dPDF}, however without the kinematic constraint
that $x_1+x_2\le 1$.

To introduce correlations, we have to create a mixed state as some
weighted sum over product states, as the GS and BA dPDFs do
in \eqref{eq:GS-dPDF} and \eqref{eq:BA-dPDF}. A
special case is a state that can be written as a classical probability
distribution over product states
\begin{equation}
  \rho^{AB} = \sum_i p_i\: \rho_i^{A} \otimes \rho_i^{B}
  \label{eq:rho_separable}
\end{equation}
with $\sum p_i = 1$, called a separable state. These states exhibit
correlations between the two subsystems, but the correlations can be
entirely explained classically\footnote{This state can easily be prepared using LOCC: use a classical random number generator to choose outcome $i$ with probability $p_i$, then locally prepare state $\rho_i^{A} \otimes \rho_i^{B}$ based on the outcome.}. Note that the subsystems can be quantum
in nature---we are only commenting on the relationships between them. 
On the other hand, a state which is not separable must have correlations 
due to quantum entanglement.

Identifying whether or not a given density matrix is separable is
known to be NP-hard~\cite{Gurvits:2003gdo,Gharibian:2008hgo}, so there
is no known universal test. However, many partial
methods~\cite{Guhne:2008qic,Horodecki:2009zz,Chruscinski:2014oca} have
been formulated for use on specific subsets of density matrices.

One standard measure of entanglement is the entanglement
entropy~\cite{Donald:2002}. Given a pure bipartite system $\rho^{AB}$
over $\mathcal{H}^{A} \otimes \mathcal{H}^{B}$, we can form a reduced
density matrix $\rho^{A}$ by only tracing over $\mathcal{H}^{B}$.  The
(von Neumann) entanglement entropy is then defined as
\begin{equation}
S_{\vn}(\rho^A) = -\tr(\rho^A\log\rho^A)\ .
\end{equation}
This definition is symmetric over $A$ and $B$; tracing over
$\mathcal{H}^{A}$ and calculating $S_{\vn}(\rho_B)$ instead will yield
the same value. If $S_{\vn}(\rho_A) > 0$, we conclude that the two
subsystems are entangled. In the next section we start from a pure
state describing the proton, but then immediately perform a partial
trace over all but two degrees of freedom (d.o.f.) to obtain a mixed
state $\rho_{x_1x_2,x_1'x_2'}$ with $S_{\vn} > 0$. Since our two-quark
density matrix is already mixed, we cannot use the entanglement
entropy to draw conclusions about the nature of correlations between the
LC momentum fractions of the quarks.

Another operation we can perform on the bipartite density matrix
$\rho^{AB}$ is to transpose only one of the subsystems $\rho^{B}$;
this is called the partial transpose and is written $\rho^{T_B}$. The
Peres-Horodecki criterion ~\cite{Peres:1996q,Horodecki:1996q} states
that if this partial transpose has any negative eigenvalues, it cannot
be separable, and hence must exhibit some quantum correlations. The
absolute value of the sum of these negative eigenvalues is called the
negativity $\mathcal{N}(\rho)$. This is a partial result since the
inverse of the criterion, that $\mathcal{N}(\rho) = 0$ implies that
$\rho$ is separable, is only true when $\dim (\mathcal{H}^{A} \otimes
\mathcal{H}^{B})$ is either $2\times 2$ or $2\times 3$.

Many other computable quantities, such as the coherent
information~\cite{Wilde:2011npi}, exist to study quantum correlations
between subsystems. Nevertheless, we shall rely on the
Peres-Horodecki criterion and entanglement negativity in our analysis
of double quark PDFs in the proton. In fact, we are mostly
interested not in quantifying the overall ``degree of
entanglement correlations'' but rather in investigating
how such correlations affect the dPDF. In this regard,
working with the partially transposed density matrix is useful
because its negative eigenvalues can be ``purged'' to construct
a nearby density matrix with negativity zero from which the
corresponding dPDF follows. This is the idea we
pursue in the next section.

\section{Double Quark Parton Distribution for the Three Quark Fock
  State} \label{sec:dPDF_qqq}

\subsection{The Three Quark Fock State}\label{sec:qqq-state}
The proton state $\ket{P}$ can be written as a coherent superposition
of all possible Fock states multiplied by their respective Fock space
amplitudes, the light-cone wavefunctions. For moderate
momentum fractions $x_i \sim 0.1$ and transverse momenta
$\vert\vec{k}_i\vert \sim \Lambda_{\text{QCD}}$, the three valence
quark state should dominate and we approximate the light-cone state
of the proton by an effective three-quark wavefunction:
\begin{equation} \label{eq:proton-lcwf}
\ket{P} = \ket{P^+, P_\perp\!\!\: = 0} = \int_{[0,1]^3} [\dd x_i] \int [\dd ^2k_i]\ \Psi_{\text{qqq}}(k_1, k_2, k_3)\ \ket{k_1;k_2;k_3}
\end{equation}
where
\begin{equation}
[\dd x_i] = \prod_{i=1\cdots 3} \frac{\dd x_i}{2x_i}\, \delta \left(1 - \sum_i x_i\right)\ ,\ \ [\dd ^2k_i] = \prod_{i=1\cdots 3} \frac{\dd ^2 k_i}{(2\pi)^3}\, (2\pi)^3\,\delta \left( \sum_i \vec{k}_i \right)\ ,
\end{equation}
and $k_i = (x_iP^+, \vec{k}_i)$ are the three-momenta of the
constituent quarks. The spatial wavefunction is symmetric
under exchange of any two quarks, $\Psi_{\text{qqq}}(k_1, k_2, k_3)
= \Psi_{\text{qqq}}(k_2, k_1, k_3)$ etc.
Since our focus is on correlations in momentum
space, we will always trace out spin-flavor and color d.o.f.\
and assume that eq.~\eqref{eq:proton-lcwf} provides an effective
description of the state in momentum space.

To construct the corresponding density matrix
we choose basis states of the form
\begin{equation}  \label{eq:basis_|alpha>}
  \ket{\alpha} = \frac{1}{16\pi^3}\, \frac{\ket{x_1,\vec{k}_1;x_2,\vec{k}_2;x_3,\vec{k}_3}}{\sqrt{x_1 x_2 x_3}}
  ~ \longrightarrow ~
  \left< P| \alpha\right> = \frac{\Psi^*_{\text{qqq}}(x_i,\vec{k}_i)}{2\sqrt{x_1 x_2 x_3}}\, \delta\!\left(1-\sum x_i\right) \delta\!\left(\sum\vec{k}_i\right)~.
\end{equation}
The density matrix for the proton state is then
\begin{equation} \label{eq:proton-dm}
  \rho_{\alpha\alpha'} = \frac{\Psi^*_{\text{qqq}}(x_i',\vec{k}_i')}{2\sqrt{x_1' x_2' x_3'}}\,
  \frac{\Psi_{\text{qqq}}(x_i,\vec{k}_i)}{2\sqrt{x_1 x_2 x_3}}
~,
\end{equation}
where the matrix ``indices'' are
$\alpha=\{x_1,\vec{k}_1,x_2,\vec{k}_2\}$,
$\alpha'=\{x_1',\vec{k}_1',x_2',\vec{k}_2' \}$; it is implied that the
r.h.s.\ is evaluated for $x_3=1-x_1-x_2$ and
$\vec{k}_3=-\vec{k}_1-\vec{k}_2$ (and similar for the primed
variables).  The trace of the density matrix is computed as
\begin{equation}
  \tr\rho = \int{\dd x_1}{\dd x_2} \int\frac{\dd^2 k_1}{16\pi^3}\frac{\dd^2 k_2}{16\pi^3}
  \, \rho_{\alpha\alpha} = 2\int[\dd x_i]\,\, \frac{1}{4}\int[\dd^2 k_i]\, \left|
\Psi_{\text{qqq}}(x_i,\vec{k}_i)
  \right|^2
  \ .
  \label{eq:tr_x1-x2}
\end{equation}

We can then construct the reduced density matrix for $x_1$ and $x_2$
by tracing over the remaining degrees of freedom, given by an
integral over the quark transverse momenta
\begin{equation}
  \rho_{x_1x_2,x_1'x_2'} = \frac{1}{4}\int[\dd^2k_i]\,\rho_{\alpha\alpha'}
~.
  \label{eq:reduced-rho_x1-x2}
\end{equation}
This reduced density matrix, which has non-zero entropy, will be
evaluated in sec.~\ref{sec:CM-variables} below.  \\

To date the exact light-cone wavefunction of the proton is
unknown, of course, as it requires
solving the QCD Hamiltonian. Here we employ a simple model due to
Brodsky and Schlumpf (BS)~\cite{Schlumpf:1992vq,Brodsky:1994fz} where
we take
\begin{equation} \label{eq:lcwf-brodsky}
\Psi_{\text{qqq}} = N\sqrt{x_1 x_2 x_3}\ e^ {-\mathcal{M}^2/2\beta^2}\ , 
\end{equation}
with $\mathcal{M}^2 = \sum (\vec k_i^2 + m_q^2)/x_i$ being the
invariant mass squared of the non-interacting three quark
system~\cite{Bakker:1979eg}. The parameters $m_q = 0.26\ \text{GeV}$
and $\beta = 0.55\ \text{GeV}$ have been tuned in
refs.~\cite{Schlumpf:1992vq,Brodsky:1994fz} to reproduce various
properties of the proton such as its electromagnetic form factors.
The normalization factor $N$ of the LCwf is determined from $\tr\rho =1$.

More involved approaches consider various model Hamiltonians with
interactions, and possibly higher Fock states such as
$|qqqg\rangle$ or $|qqqq\bar{q}\rangle$, and the quark model LCwf are then obtained numerically.
Ref.~\cite{Qian:2024fqf} has recently studied quark and gluon
entanglement in the proton in this way. However, the wavefunctions
and density matrices are then available only numerically. For this
first exploration of quantum correlations associated with the
entanglement negativity measure we consider the simple BS model above.

\subsection{Purging Entanglement Negativity with the PEN transformation} \label{sec:PEN}

The PEN (Purge Entanglement Negativity) algorithm, introduced and detailed in
ref.~\cite{Dumitru:2023fih}, can be used to remove the negativity of a
density matrix. Namely, it takes a density matrix $\rho$ and produces
a new density matrix $\rho'$ such that $\mathcal{N}(\rho') = 0$ by
replacing all negative eigenvalues of $\rho^{T_2}$ with
zeroes\footnote{The resulting $\rho'$ is, indeed, a valid density
matrix in that it has unit trace, and is hermitian and positive
semi-definite.}. In short, after diagonalizing $\rho^{T_2}$ the negative
eigenvalues are replaced by 0; we then undo the unitary transformation
to the eigenbasis of $\rho^{T_2}$ and the partial transposition, and finally
rescale the matrix by $1/(1+\mathcal{N}(\rho))$ so that $\tr \rho' = 1$.
Note that $\rho'$ is not guaranteed to be separable,
except for small Hilbert spaces where the Peres-Horodecki criterion is
sufficient to establish separability. However, since the negative
eigenvalues of the partial transpose of $\rho$ are necessarily
associated with quantum correlations, these specific correlations must
be removed when the associated eigenvalue is sent to zero.

In some low-dimensional cases such as Bell states\footnote{A different
example is given in appendix~B of ref.~\cite{Dumitru:2023fih}. There,
a $\nc^2 \times \nc^2$ dimensional anti-symmetric color space state is
considered where $\rho^{T_2}$ also exhibits a single negative, and a
fully degenerate spectrum of positive eigenvalues. It is shown that
the application of PEN results in the minimal admixture of the
identity so as to achieve a separable Werner state.}, where
$\rho^{T_2}$ features a single non-degenerate negative eigenvalue,
$\rho'$ is the closest separable state to $\rho$, i.e.\ a separable
Werner state~\cite{Dahl:2006}. If $\rho$ is separable to begin with
then the PEN transformation reduces to the identity and $\rho' =
\rho$.

\subsection{Leveraging Center of Mass Variables}
\label{sec:CM-variables}

In this section we construct the reduced density matrix for the LC
momentum fraction degrees of freedom of two quarks.  The transverse
momentum integrals over the BS wavefunction in
eq.~\eqref{eq:reduced-rho_x1-x2} can be done by hand, yielding
\begin{equation} \label{eq:BS-dPDF-x1x2}
\rho_{x_1x_2,x_1'x_2'} \sim \frac{\exp\left[-\frac{m_q^2}{2\beta^2}(A_1 + A_2 + A_3) \right]}{A_1A_2 + A_1A_3 + A_2A_3} 
\end{equation}
with $A_i = \frac{1}{x_i} + \frac{1}{x_i'}$. Recall that
$x_3$ is not an independent variable, and should
be read as shorthand for $1-x_1-x_2$ (likewise for
$x_3'$). We now have a mixed density matrix with two subsystems for
the two valence quark LC momentum fractions. 

The denominator $A_1A_2 + A_1A_3 + A_2A_3$ does not seem readily
factorizable into a polynomial in $x_1$ times a polynomial in $x_2$ on
the diagonal, suggesting that this is not a classical dPDF i.e.\ it
cannot be written as a classical mixture of product states. Our goal
is to confirm this guess by numerically exhibiting the presence of
quantum correlations in the dPDF. If we go about this the naive way,
using the PEN algorithm to erase the negativity of
$\rho_{x_1x_2,x_1'x_2'}$, the resulting density matrix does not
respect the kinematic $x_1+x_2\le 1$ constraint of the problem and
gives an unphysical dDPF. The simple reason for this is that given
$x_1 + x_2 \leq 1$ and $x_1' + x_2' \leq 1$, partial
transposition (swapping $x_2$ and $x_2'$) does not necessarily respect
these constraints.

The solution is to use center-of-mass (c.o.m.) coordinates that
decouple relative and c.o.m.\ motion, and factorize the constraints of
the problem~\cite{Bakker:1979eg}. By finding the dPDF in the
c.o.m.\ variables, using the PEN algorithm on the c.o.m.\ density
matrix, then changing back to our original variables $x_1$ and $x_2$,
we remove negativity while preserving the momentum constraint.

A set of suitable relative light-cone coordinates for three particles
constrained to $x_1+x_2+x_3=1$ is~\cite{Bakker:1979eg}
\begin{align}
  \xi = \frac{x_1}{x_1+x_2}\ ,\ \eta = x_1 + x_2\ ,& \label{eq:xietadef}
\end{align}
with $\xi$, $\eta \in [0,1]$.  In terms of these variables,
\begin{equation}
2[\dd x_i]\,\, \frac{1}{4}[\dd^2k_i] =
\frac{\dd\xi\, \dd\eta}{2\xi(1-\xi)\, 2\eta(1-\eta)}\,\,
\frac{1}{4}[\dd^2k_i]~,
\label{eq:measure_x1-x2-->xi-eta}
\end{equation}
and
\begin{equation}
  \rho_{\xi \eta, \xi' \eta'} = \frac{1}{4}\int[\dd^2 k_i]\,
  \frac{\Psi^*_{\text{qqq}}(\xi'\eta',\eta'(1-\xi'),1-\eta',
    \vec{k}_1,\vec{k}_2,\vec{k}_3)}
  {\sqrt{4\eta'(1-\eta')\xi'(1-\xi')}}\,
  \frac{\Psi_{\text{qqq}}(\xi\eta,\eta(1-\xi),1-\eta,
    \vec{k}_1,\vec{k}_2,\vec{k}_3)}
  {\sqrt{4\eta(1-\eta)\xi(1-\xi)}}
  \label{eq:rho-red_xi-eta}
\end{equation}
is the reduced density matrix over the remaining $\xi$, $\eta$ degrees
of freedom.  It is normalized such that
\begin{equation}
  1 = \int \dd\xi\dd\eta\, \rho_{\xi \eta, \xi \eta}\ .
  \label{eq:xietatrace}
\end{equation}

\begin{figure}[ht]
\centering
\includegraphics[width=0.6\textwidth]{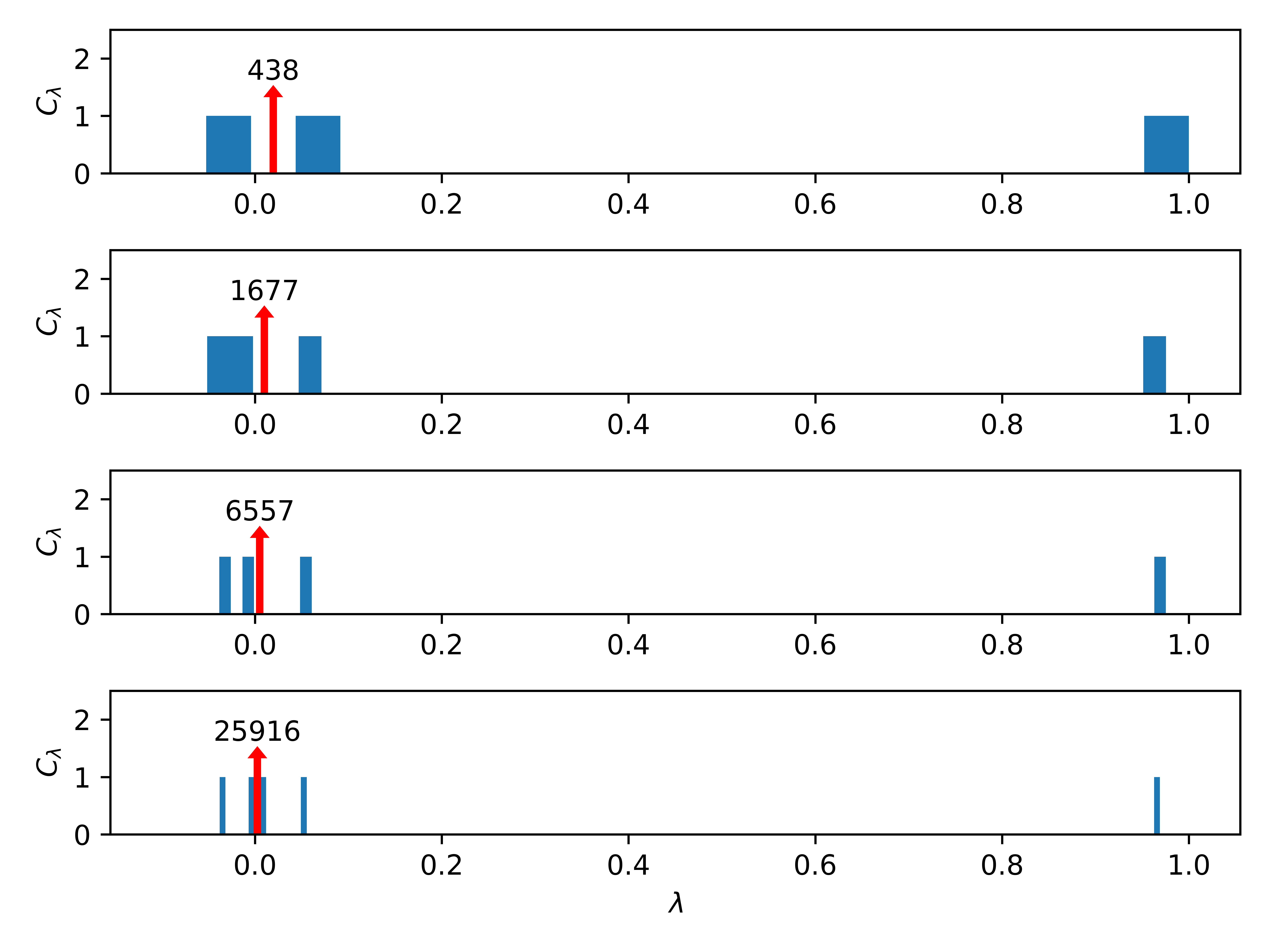}
\vspace*{-.3cm}
\caption{\label{fig:ppt-eigen} Eigenvalue spectrum of the partial
  transpose $\tilde\rho^{T_2}$ for $\frac{1}{\Delta\xi} =
  \frac{1}{\Delta\eta} =$ 20, 40, 80, 160. The red arrows at the
  origin signify $\delta(\lambda)$ peaks in the continuum limit. We
  have chosen to show these peaks as arrows labeled with their correct
  height, rather than plotting these peaks and obscuring the nonzero
  eigenvalues.}
\end{figure}

So far $\xi$ and $\eta$ are continuous variables on $[0,1]$. In order
to evaluate $\rho_{\xi \eta, \xi' \eta' }$ on a computer and analyze
it with information theoretical techniques, we need a finite
dimensional matrix.  This is achieved by discretizing the unit interval
into a finite number of bins of size $\Delta\xi$ and $\Delta\eta$.
The discretized density matrix is
\begin{equation} \label{eq:normrhoxieta}
  \tilde\rho_{\xi \eta, \xi' \eta' } =
  {\Delta\xi\,\Delta\eta}\ \rho_{\xi \eta, \xi' \eta'}\ . 
\end{equation}
We can then construct the partial transpose $\tilde\rho^{T_2}$ and
numerically calculate the negativity. For the BS light-cone
wavefunction, we employed $\Delta\xi = \Delta\eta$ bins of width
$\frac{1}{N}$ for $N = 20,40,80,160$ and found that
$\mathcal{N}(\tilde\rho_{\xi \eta, \xi' \eta' })$ converges to
$\sim\!0.035$. This is because once the variation of
$\tilde\rho^{T_2}$ across a bin is small a further decrease of the
bin width only increases the dimension of the null space of the
matrix. That is, as the number $N$ of bins increases,
the eigenvalue density of $\tilde\rho^{T_2}$ asymptotically approaches
the form
\begin{equation}
\frac{\dd N_\lambda}{\dd\lambda} = \left((N+1)^2 - \sum_{i=1}^{n}C_i\right)\delta(\lambda) + \sum_{i=1}^{n}C_i\,\delta(\lambda-\lambda_i)
\end{equation}
where $C_i$ is the multiplicity of the nonzero eigenvalue
$\lambda_i$. This can be seen in fig.~\ref{fig:ppt-eigen}. 
There are $N_\lambda=(N+1)^2$ total eigenvalues of the
partial transpose, but all but a few of them remain in a delta
function peak at zero. The nonzero eigenvalues do not drift as $N$
changes, so increasing $N$ only increases the resolution for these
values. One notable nonzero eigenvalue is $-0.03$ which provides most
of the contribution to the negativity of $\tilde\rho$.

Once we use the PEN algorithm to purge the negative eigenvalues of the
partial transpose, we can return to (discrete) $x_1$-$x_2$ space for
example, to compute single and double PDFs as functions of $x_1, x_2$
as described below.

For the BS light-cone wavefunction, we find the negativity of
$\rho_{\xi \eta, \xi' \eta' }$ to be $\sim\!0.035$. This may be
interpreted in the sense that quantum correlations are weak. However,
we already mentioned at the end of sec.~\ref{sec:Detection_qC}
that we are mostly interested not in the overall (averaged over
momentum fractions) ``degree of entanglement'', but in how
the presence of quantum correlations affects the dPDF.
In the following section we shall see that purging the negativity of
the density matrix can result in substantial modification of the dPDF,
at least in specific regions of $x_1$-$x_2$ space.

\subsection{Numerical Results for the Quark dPDF}
\label{sec:dPDF-results-LO}

We can compare the structure of the dPDF, both pre- and post-PEN, 
to models such as the GS model \eqref{eq:GS-dPDF} 
and the BA model \eqref{eq:BA-dPDF} by plotting 
the ratio of the dPDF to the product of two single-quark PDFs,
\begin{equation}  \label{eq:C-x1-x2}
C(x_1,x_2) = \frac{f_{qq}(x_1,x_2)}{f_q(x_1)\, f_q(x_2)}~.
\end{equation}
If the dPDF is simply a product of two single-quark PDFs without any
correlations, then this ratio is 1. The dPDF is given by the diagonal
elements of the reduced density matrix defined above in
eqs.~(\ref{eq:proton-dm}, \ref{eq:reduced-rho_x1-x2}),
\begin{equation}
  f_{qq}(x_1,x_2) = \rho_{x_1 x_2,x_1 x_2}~.
  \label{eq:dPDF-red_rho}
\end{equation}
This relation can be confirmed by direct computation of the
matrix element in the proton state~\eqref{eq:proton-lcwf} of
the operator
\begin{equation}
  \frac{\pi\, P^+}{(2\pi)^3}\,
  \int\dd^2 z \int \dd z_1^-\dd z_2^-\dd z_3^-\,
  e^{-i x_2 P^+ (z_1^--z_2^-) -i x_1P^+ z_3^-}
  O(z_1^-+\vec z,z_2^-+\vec z)\, O(z_3^-,0)
\end{equation}
with $O(z,y) = \bar{q}(z) \gamma^+ q(y)$, and $q(z)$ the quark field
operator at point $z$. (The notation $z^-+\vec z$ refers to a vector
with LC minus component $z^-$, and transverse components $\vec z$ which
represent the transverse spatial separation of the quarks.)
In
sec.~\ref{sec:rho_diagonal-DGLAP} below we verify that
$f_{qq}(x_1,x_2)$ satisfies the dPDF Dokshitzer-Gribov-Lipatov-Altarelli-Parisi (DGLAP) equation upon emission of a
collinear gluon.

The single quark PDF is given by\footnote{Eq.~\eqref{eq:f_q(x1)} can
be confirmed by computing the Dirac electromagnetic form factor which
in the limit of vanishing momentum transfer reduces to $\int \dd x\,
f_q(x)=F_1(0)=1$.} the integral of the dPDF
over $x_2$ from 0 to $1-x_1$,
\begin{equation}
  f_q(x_1)=\int\limits_0^{1-x_1}\dd x_2\,
  f_{qq}(x_1,x_2)~.  \label{eq:f_q(x1)}
\end{equation}
The dPDF momentum sum rule~\cite{Gaunt:2009re} is satisfied\footnote{That is,
$$\int\dd x_2\, x_2 f_{qq}(x_1,x_2) +
\int\dd x_3\,x_3 f_{qq}(x_1,x_3) = (1-x_1)f_q(x_1)~.$$} thanks to
$f_{qq}(x_1,x_2)=f_{qq}(x_1,1-x_1-x_2)$. We also confirm that
$3\int\dd x_1\, x_1\, f_q(x_1) = 2\int[\dd x_i]\, (x_1+x_2+x_3)\,
\frac{1}{4}\int[\dd^2k_i]\,
|\Psi_{\text{qqq}}(x_1,x_2,x_3,\vec{k}_1,\vec{k}_2,\vec{k}_3)|^2 =
\tr\rho=1$.

\begin{figure}[htb]
	\includegraphics[width=0.48\textwidth]{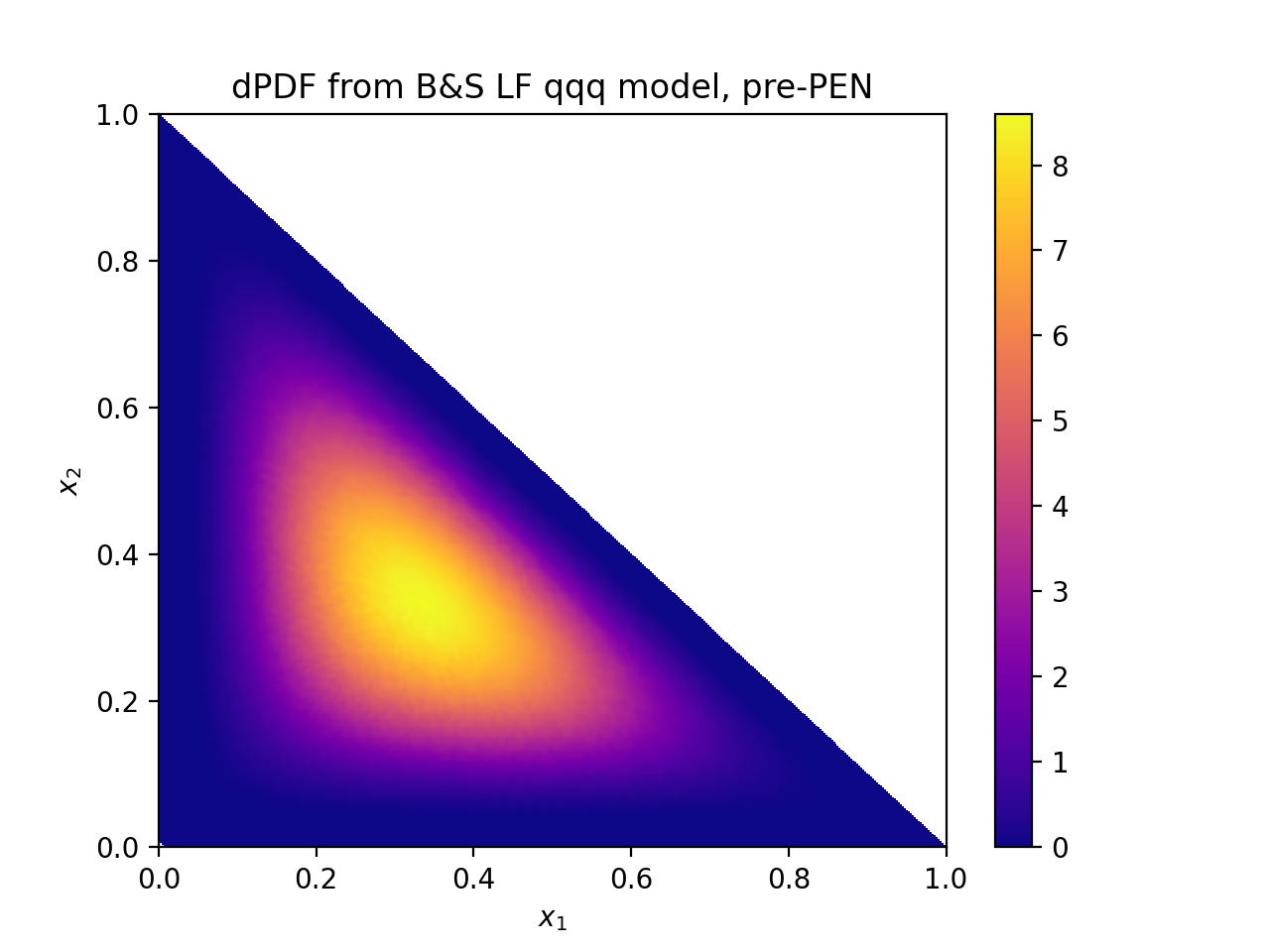}
	\includegraphics[width=0.48\textwidth]{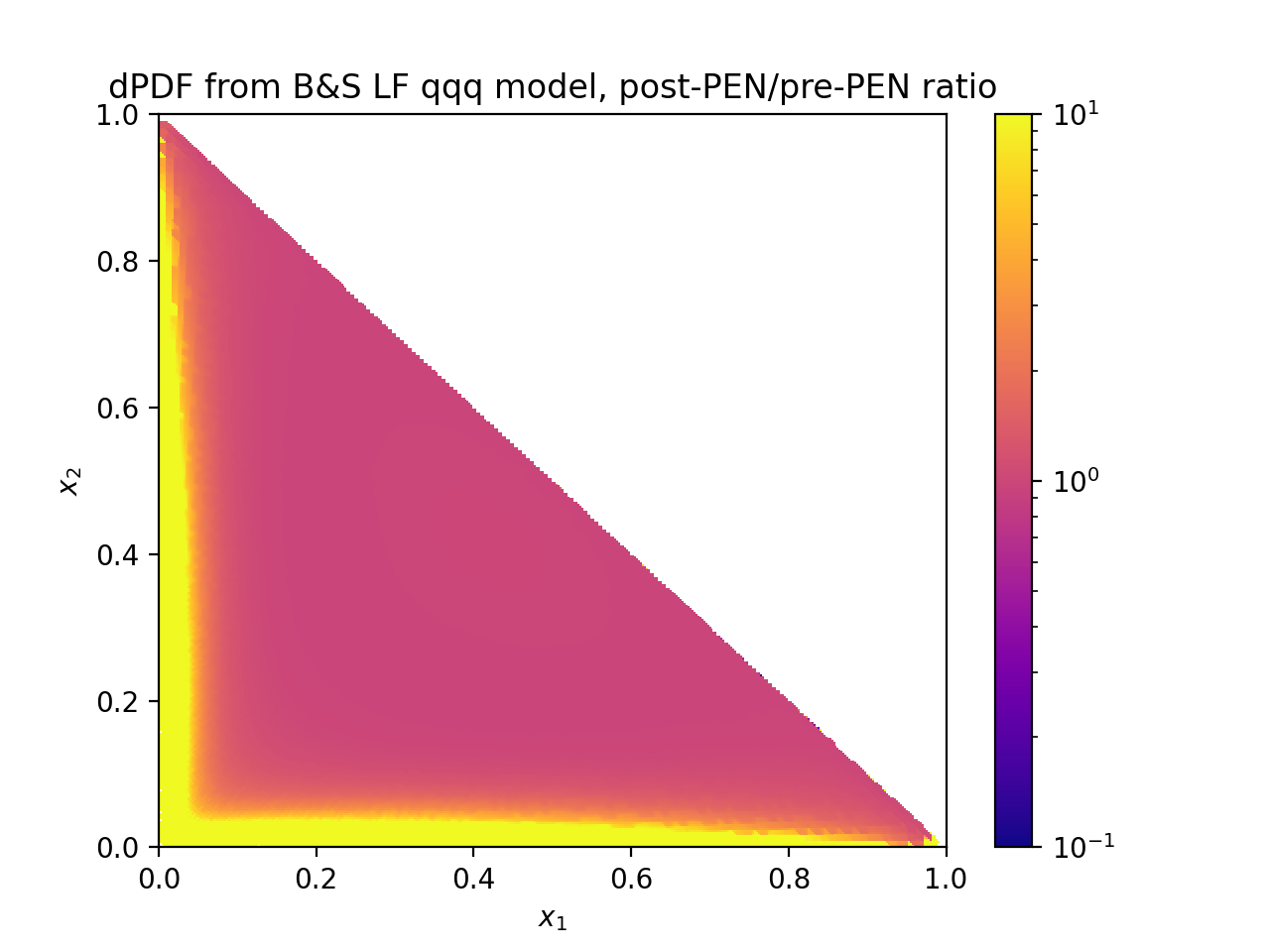}
	\vspace*{-.3cm}
	\caption{Left: the double quark PDF of the LC quark model (see text)
		in the $x_1$-$x_2$ plane. Right: the ratio of the double quark PDF
		after removal of the entanglement negativity to the original dPDF
		shown on the left.}
	\label{fig:dPDF-x1-x2}
\end{figure}
In fig.~\ref{fig:dPDF-x1-x2}(left) we show the double
quark PDF $f_{qq}(x_1,x_2)$ of the LC quark model described in
sec.~\ref{sec:qqq-state}. As expected, it peaks around $x_1 \sim x_2
\sim 1/3$ with tails extending to smaller $x_i$. The panel on the
right shows the dPDF after removal of the entanglement negativity of
the density matrix (via the PEN transformation), divided by the
original dPDF. In much of the allowed range of $x_1, x_2$ the PEN
transformation has affected the dPDF by a factor of order 1, hence
PEN is close to the identity map over a large portion of the domain.
For $x_1$ or $x_2$ less than approximately 0.1 the dPDF has
changed by a factor of 10 or more. Of course, in the region of small
$x_i$, and for $Q^2$ greater than the confinement scale, one expects
sea partons to dominate. In sec.~\ref{sec:rho_evol} below we shall
study the effects on the quark dPDF due to the perturbative emission
of a gluon.

\begin{figure}[htb]
  \includegraphics[width=0.48\textwidth]{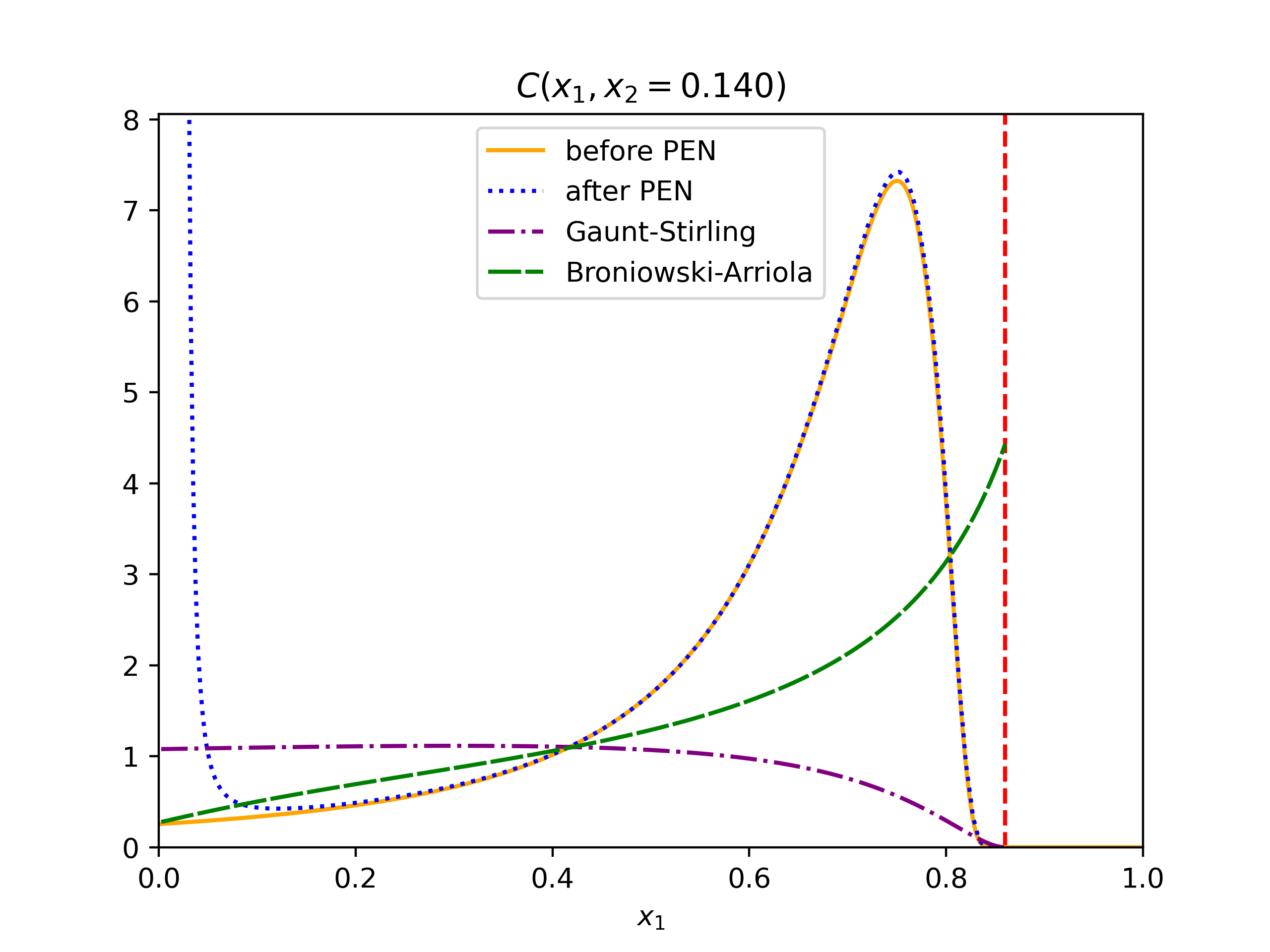}
  \includegraphics[width=0.48\textwidth]{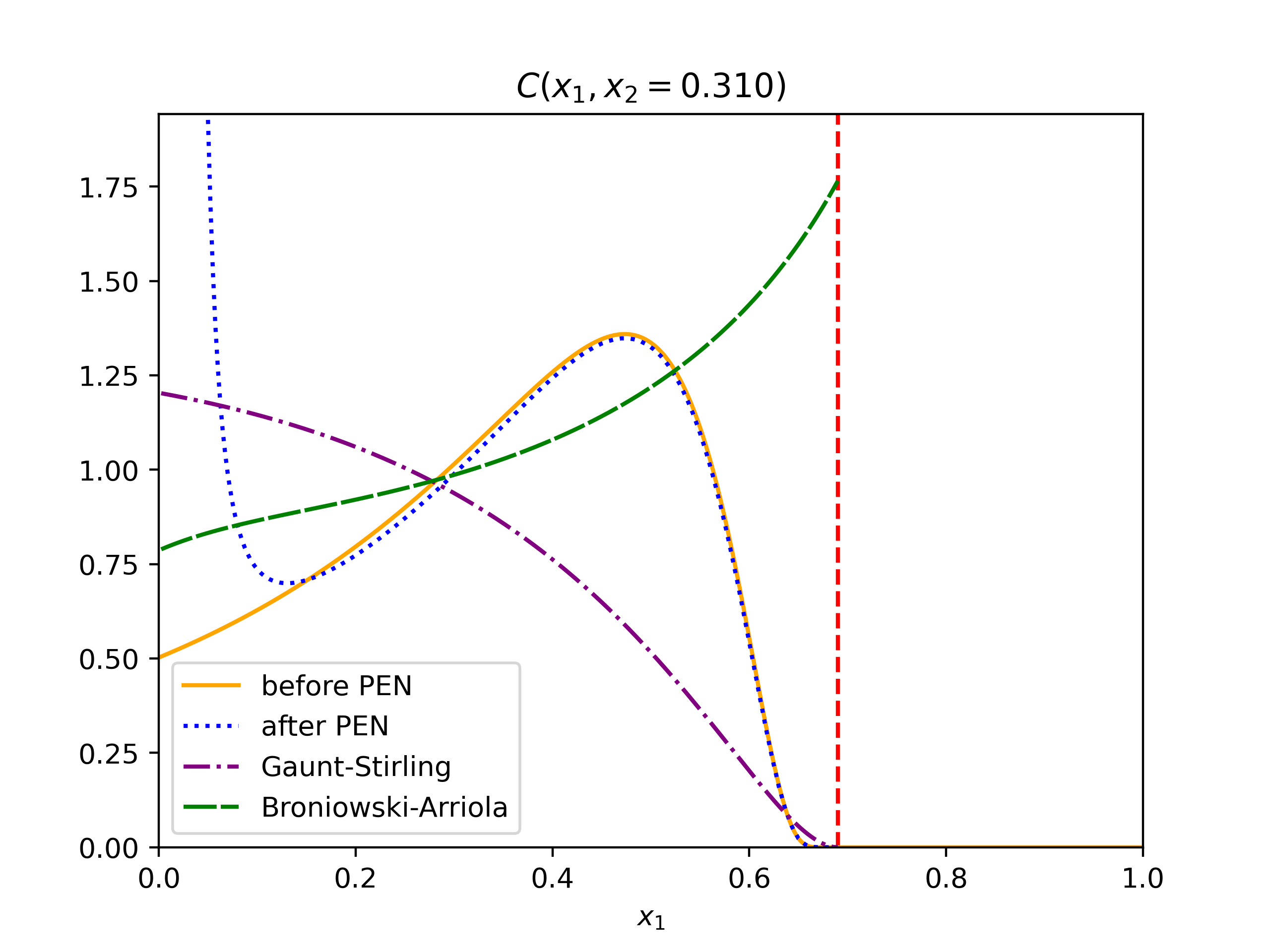}
  \includegraphics[width=0.48\textwidth]{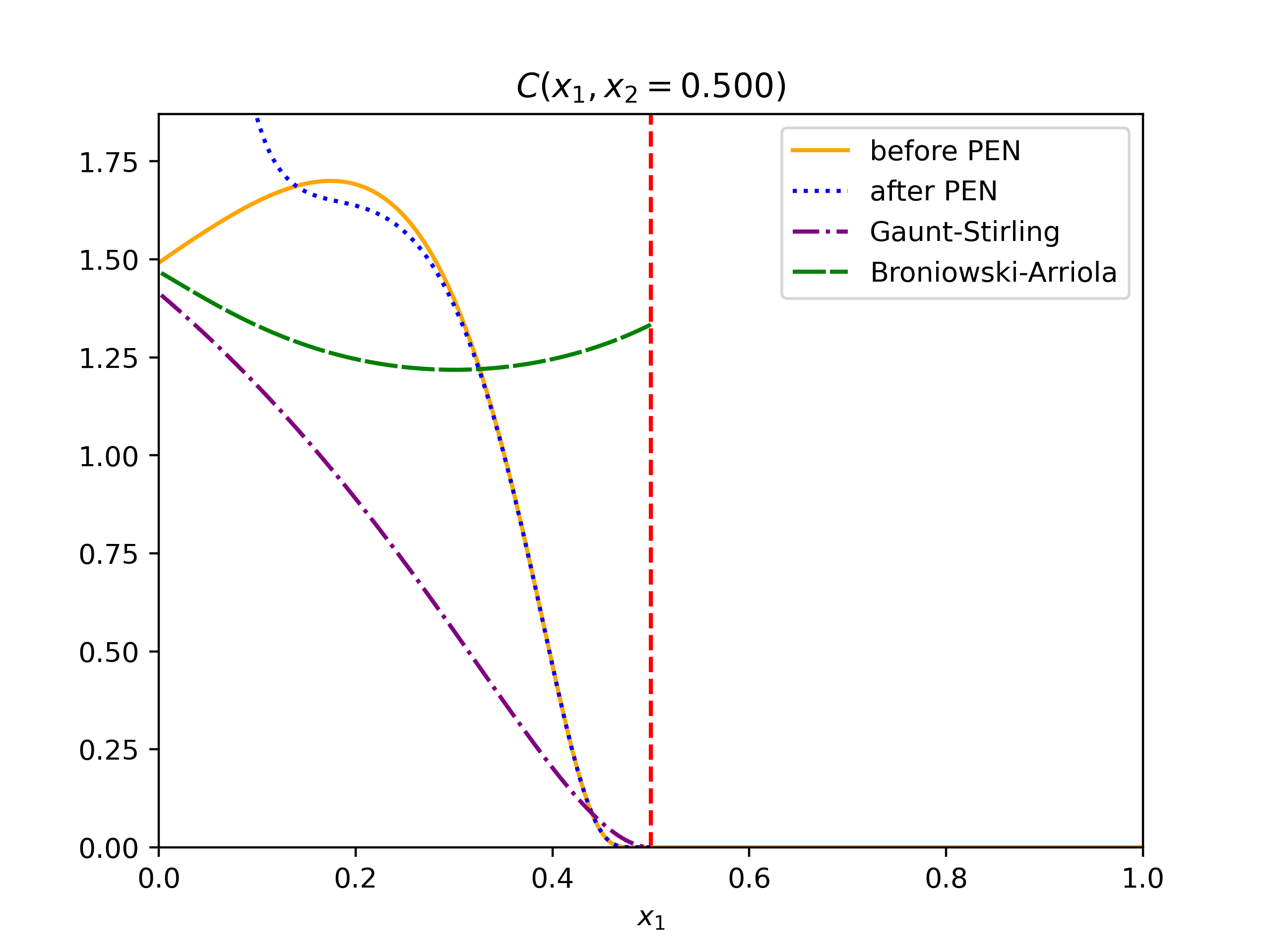}
  \includegraphics[width=0.48\textwidth]{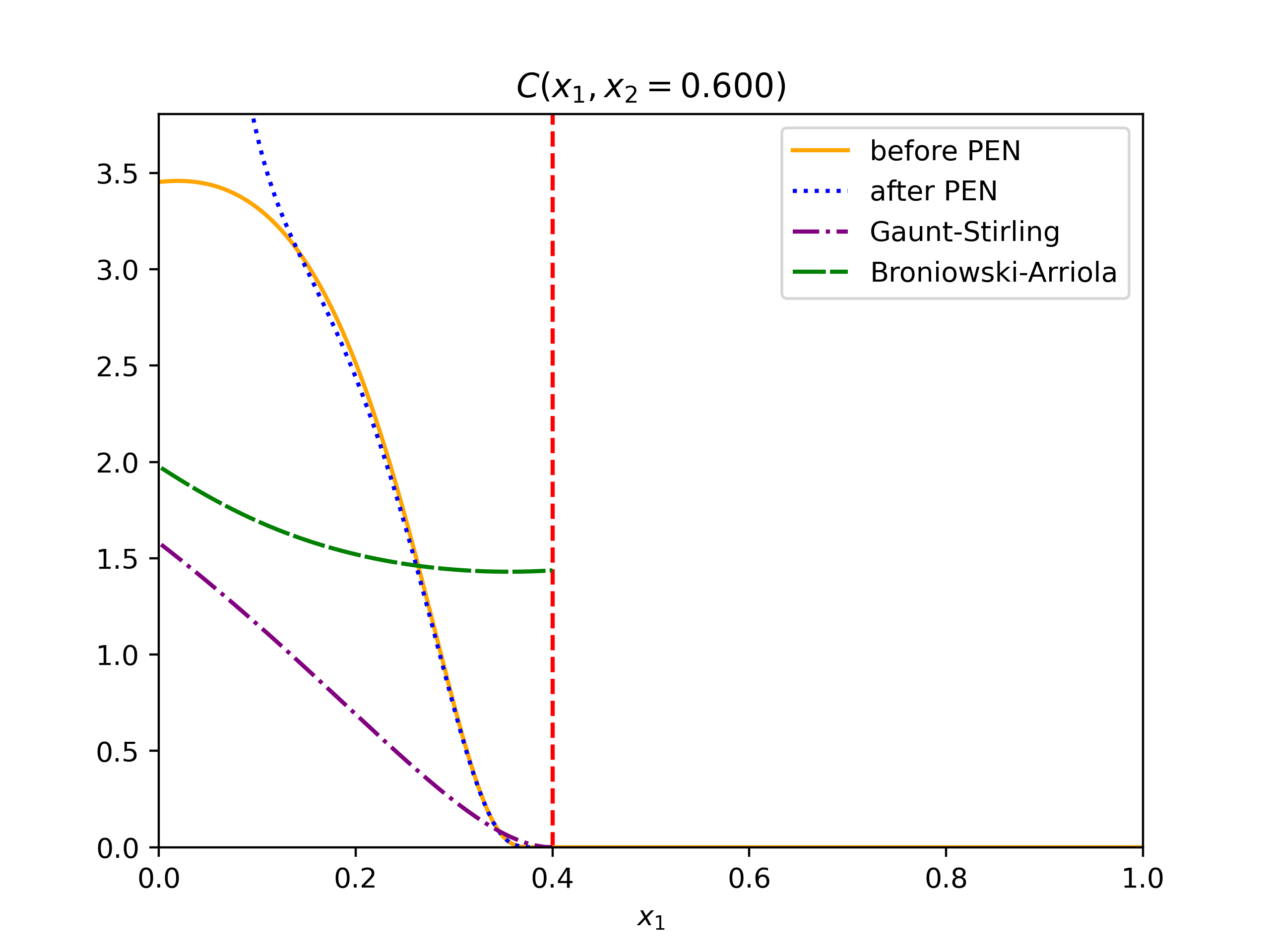}
  \vspace*{-.3cm}
  \caption{\label{fig:pen-dpdf-ratio} Some selected slices of constant
    $x_2$ for the ratio of the dPDF to the product of two PDFs. The
    vertical dashed line represents the kinematic boundary where
    $x_1+x_2=1$.  The dash-dotted and dashed curves correspond to the
    GS~\cite{Gaunt:2009re} and BA~\cite{Broniowski:2013xba} models
    given in eqs.~(\ref{eq:GS-dPDF},\ref{eq:BA-dPDF}),
    respectively. The solid and dotted curves refer to the dPDF
    obtained here from the Brodsky-Schlumpf model
    LCwf~\cite{Schlumpf:1992vq,Brodsky:1994fz} before and after the
    removal of quantum correlations via the PEN algorithm.}
  \label{fig:dPDF-slices}
\end{figure}
Our main interest is in the nature of correlations encoded in the
dPDF. Hence, in fig.~\ref{fig:dPDF-slices} we plot the
correlation factor $C(x_1,x_2)$, eq.~(\ref{eq:C-x1-x2}), as a
function of $x_1$ for various $x_2$. For small $x_2 = 0.14$, the GS model predicts weak
correlations up to $x_1\approx 0.5$ which then turn into an
anti-correlation as the boundary of phase space is approached.  The
correlation measure obtained from the BS wavefunction is close to the
BA model for a large portion of phase space from $x_1 \sim 0.05$ to
$\sim 0.5$; the anti-correlation at small $x_1$, $C(x_1,x_2)<1$,
weakens with increasing $x_1$.  The BS wavefunction predicts a much
stronger correlation for large $x_1\simeq0.7$ but this extreme part of
phase space is less relevant in practice. {\em Removal} of the quantum
correlations associated with negativity (by application of PEN)
does not affect the dPDF much when $x_1 \gtrsim 0.1$ but it does
lead
to a strong correlation peak at small $x_1 < 0.1$. Hence,
assuming that the GS, BA, and the BS LCwf models are close to
the real dPDF of QCD we conclude that the LC quark model
wavefunction requires the presence of subsystem
quantum correlations so as not to strongly overpredict the dPDF
at small $x_2$, and yet smaller $x_1 < x_2$.

For $x_2 \simeq 0.3$ we still see fair agreement between the dPDF from
the BS wavefunction and the BA model from $x_1 \sim 0.1$ to $\sim
0.5$.  Here, the two qualitatively differ from the GS curve which
shows a monotonic decrease and a suppressed dPDF at $x_1 \ge 0.3$.
However, the LC quark model and the GS model both predict a strong
suppression of the dPDF towards the edge of phase space, unlike the BA
model. We also
again observe the appearance of a strong enhancement of the dPDF of the
LC quark model at
small $x_1 \le 0.1$ when the quantum correlations (associated with the
negativity measure) of the two quark density matrix are removed.

For the panels in the first row of fig.~\ref{fig:dPDF-slices} we
may identify the following structure in the BS model. There exists a
{\it quantum correlated} regime at small $x_1 < x_q^*$ where the
positive slope of $C(x_1,x_2)$ is due to quantum correlations; this
regime extends up to about $x_q^* \approx 0.1$--$0.15$ where the dotted and
solid lines merge.  Beyond $x_q^*$, correlations are classical as the
dotted and solid lines are on top of each other. However, we can
subdivide this further into a {\it classically correlated} regime up
to $x_c^{*}$, which is the point where the curves reach their
maximum. In the classically correlated regime between $x_q^*$ and
$x_c^*$, $C(x_1,x_2)$ again exhibits a positive slope.  Lastly,
beyond $x_c^*$ we have the {\it cutoff} regime where the dPDF drops
to 0 very rapidly, faster than the sPDF $f_q(x_1)$, and the slope of
$C(x_1,x_2)$ is negative. This arises due to the increasing importance
of the momentum constraint.

Proceeding to greater $x_2 = 0.5$ we again observe a strong
enhancement of the dPDF at small $x_1 \lesssim 0.2$ due to PEN.
This {\it quantum correlated} regime then transitions right away into the
{\it cutoff} regime, i.e.\ the intermediate {\it classically
  correlated} regime for $x_q^* < x_1 < x_c^*$ has been ``squeezed
out'' by the merging {\it quantum correlated} and {\it cutoff}
regimes.

For $x_2 = 0.6$, the pre- and post-PEN curves exhibit similar
behavior. Small phase space suppresses quantum correlations in that
both curves exhibit a downward trend. However, the presence of
quantum correlations again prevents a stronger enhancement of the dPDF
below $x_1 \sim 0.1$. This is maintained as $x_2$ heads towards the
upper extreme of phase space.

Summarizing this section, we find that for rather asymmetric momentum
fractions $x_1$ and $x_2$, and well below the kinematic boundary
$x_1+x_2=1$, the BS model LCwf does exhibit significant quantum
correlations associated with the negativity measure. On the other
hand, for $x_1 \simeq x_2$ or large $x_1+x_2$, while the dPDF can be
substantially different from the product of single quark PDFs,
nevertheless these correlations are not associated with entanglement
negativity.

\section{Evolution to Higher Scales} \label{sec:rho_evol}

So far we have considered the correlations of the dPDF at a low
resolution scale $Q_0^2$ where a LC quark model may apply.  Naturally,
one may be interested in how Fig.~\ref{fig:dPDF-slices} changes as we
evolve to higher scales. The DGLAP evolution of dPDFs
to high $Q^2$ has been studied extensively in the literature, see
e.g.~\cite{Korotkikh:2004bz,Gaunt:2009re,Blok:2010ge,Diehl:2011yj,Diehl:2017wew,Broniowski:2013xba,Golec-Biernat:2014bva,Golec-Biernat:2015aza}. However,
analysis of classical vs.\ quantum correlations using negativity and
PEN requires knowledge of the entire density matrix, not just the
diagonal elements, so we must evolve the entire density matrix.
Hence, we require an extension of the dPDF DGLAP equations to
off-diagonal density matrix elements. Here we consider only the first step of scale
evolution where one of the three valence quarks splits into a quark
and a gluon. We consider diagrams which exhibit a collinear
divergence.  In this section we will use expressions derived in sec.~3
of ~\cite{Dumitru:2022tud}, and in~\cite{Dumitru:2020gla} where the
quark wavefunction renormalization factor (virtual corrections) and
the Fock space amplitude in the LC gauge of the $|qqqg\rangle$ state (real
emissions) have been determined.

\subsection{Corrections to the density matrix due to one gluon
emission}

\begin{figure}[htb]
	\centering
  \includegraphics[width=0.2\textwidth]{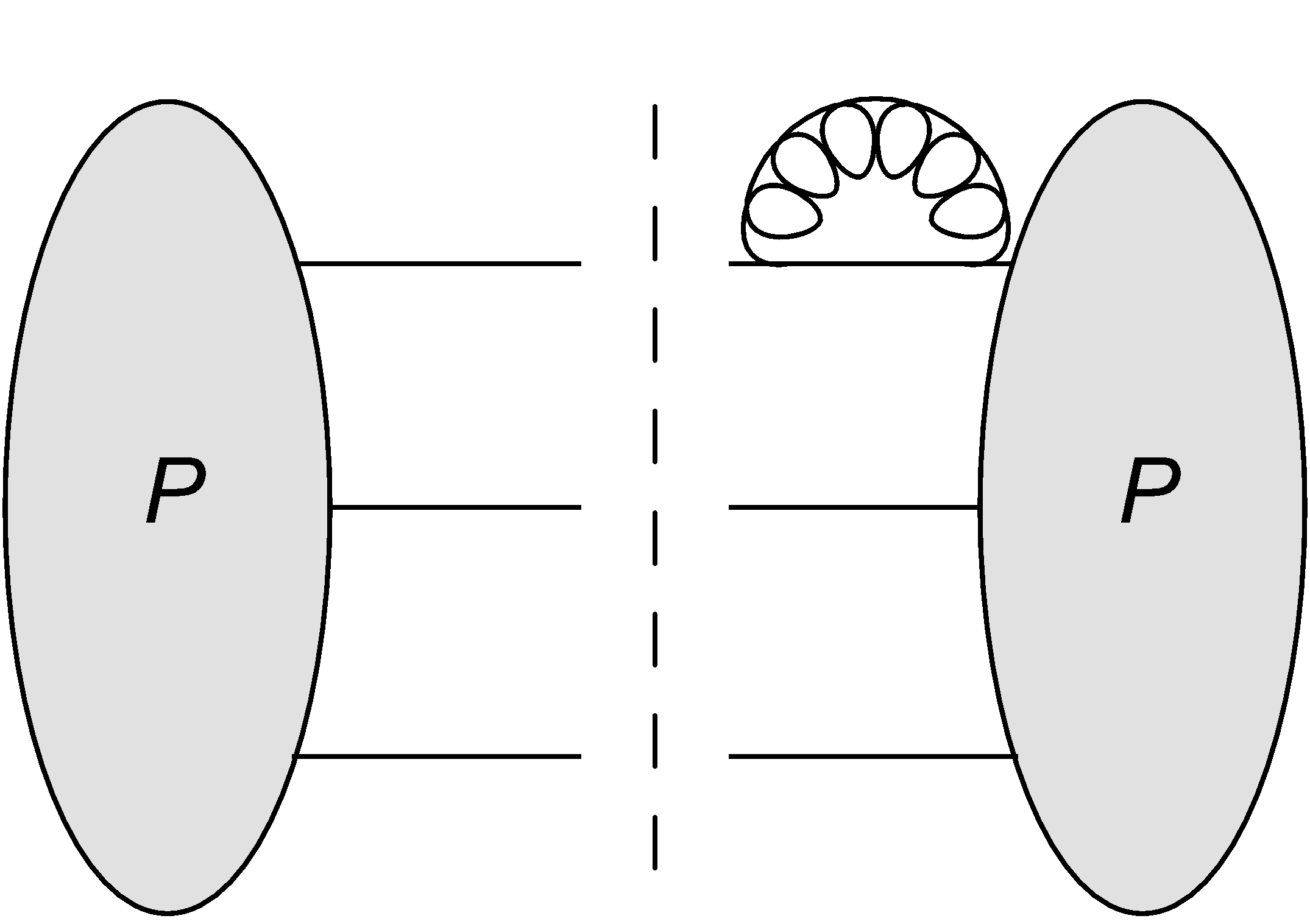}\hspace{3em}
  \includegraphics[width=0.2\textwidth]{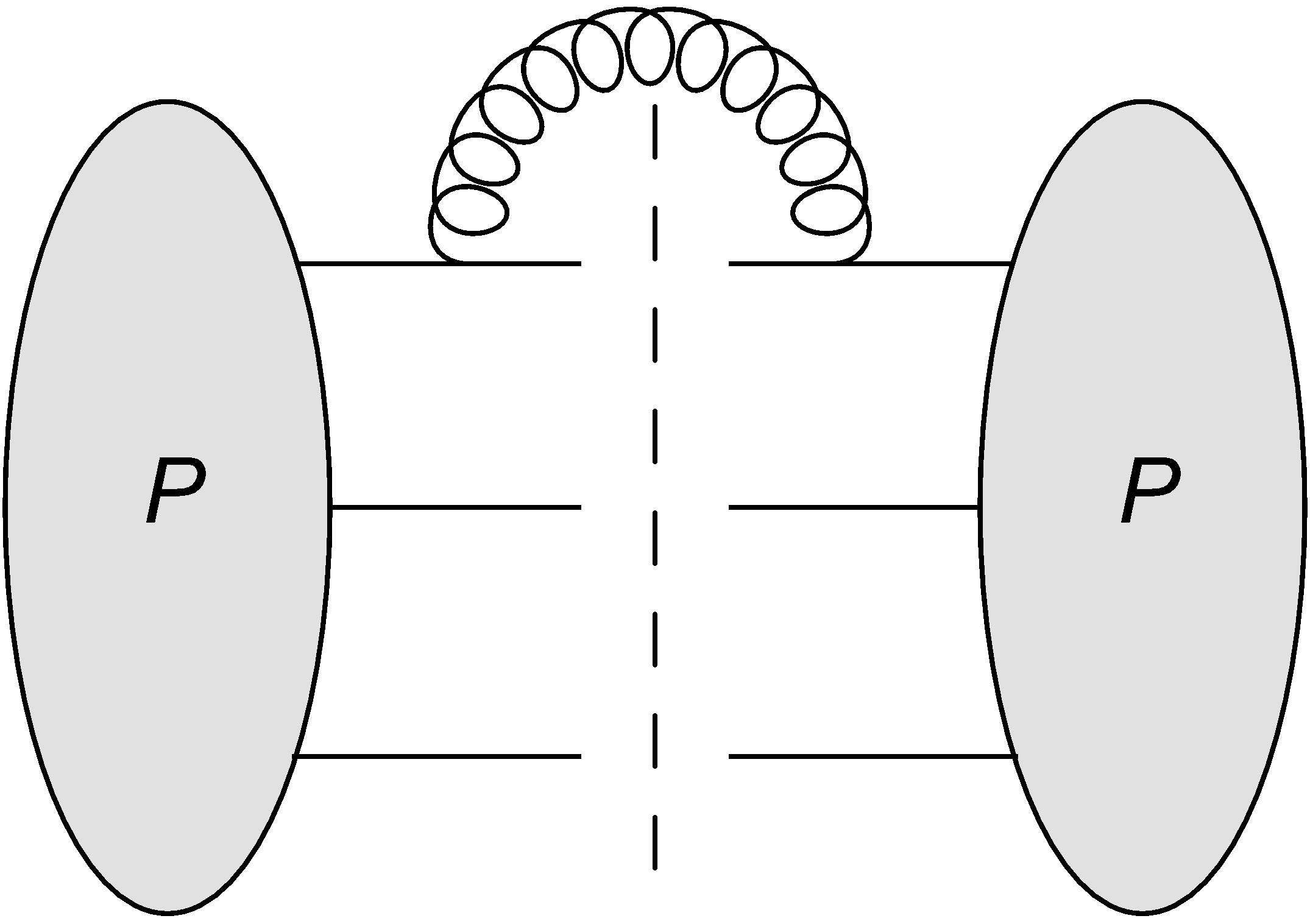}
  \vspace*{-.2cm}
  \caption{\label{fig:dglap-diag-ex} Two diagrams representing,
    respectively, the emission and absorption of a gluon by quark 1,
    and the exchange of a gluon between quarks 1' and 1. The dashed
    line represents the insertion of two-quark basis states. There are
    six diagrams of the first type and three of the second type.}
  \label{fig:dglap-diagrams-ex}
\end{figure}

Depending on where we insert our basis states, $\rho_{\alpha\alpha'}=
\langle P|\alpha\rangle\, \langle\alpha'|P\rangle$, in fig.~\ref{fig:dglap-diagrams-ex}, we have to
consider either a gluon emitted and absorbed by the same quark in
$\ket{P}$ or $\bra{P}$, i.e.\ a virtual correction, or a ``real
emission'' diagram where a gluon is exchanged by quark $i$ in
$\ket{P}$ and the corresponding quark in $\bra{P}$.

The first case gives the quark wavefunction renormalization factor
$Z_q^{1/2}=(1-C_q)^{1/2} \simeq 1-C_q/2$ via
\begin{align}
  C_q\left(x_i; x,\frac{M_{\text{UV}}}{m_{\text{col}}}\right) &=
  \frac{2g^2C_F\nc}{3} \int_x^1 \frac{\dd x_g}{x_g} \intddkg\
  \Theta(x_i-x_g) \left[ 1 + \left(1 - z\right)^2\right]
  \left[\frac{1}{k_g^2 + \Delta^2} - \frac{1}{k_g^2+\Lambda^2}\right] \\
  &= \frac{\alpha_s}{\pi} C_F \, \log\frac{M_{\text{UV}}^2}{m_{\text{col}}^2}
  \left[\log\frac{x_i}{x} - \frac{(3x_i-x)(x_i-x)}{4x_i^2}
    \right]\, \Theta(x_i-x)~.
  \label{eq:cqdef}
\end{align}
Here $z = x_g/x_i$, and $\Delta = zm_{\text{col}}$ where
$m_{\text{col}}$ denotes a quark mass regulator for the collinear
singularity. $\nc=3$ the number of colors in QCD, $C_F =
(\nc^2-1)/2\nc = 4/3$ the quadratic Casimir of the fundamental
representation, and $\alpha_s = g^2/4\pi$ the coupling. The factor
$C_F[1 + \left(1 - z\right)^2]$ represents, of course, the splitting
function $zP_{g\leftarrow q}(z)$.

We have regularized the UV divergence by introducing a cutoff with
$\Lambda = zM_{\text{UV}}$, and the soft divergence by a cutoff $x$ on
the LC momentum fraction of the gluon. The above expression
corresponds to the first step of evolution taking us from the scale
$m_{\text{col}}$ to $M_{\text{UV}}$.  Since we consider just one
single collinear gluon emission we must require that $C_q(x_i) \ll 1$.

The wavefunction renormalization factor arises for each of
the six quark lines:
\begin{align}
  \Psi^*_{\text{qqq}}(\xi',\eta',\vec{k}_i')\,
  \Psi_{\text{qqq}}(\xi,\eta,\vec{k}_i)
  &\rightarrow \Psi^*_{\text{qqq}}(\xi',\eta',\vec{k}_i')\,
  \Psi_{\text{qqq}}(\xi,\eta,\vec{k}_i)
  \left[1 -
    \frac{1}{2}C_q\left(x_1; x,\frac{M_{\text{UV}}}{m_{\text{col}}}\right) -
    \frac{1}{2}C_q\left(x_1'; x,\frac{M_{\text{UV}}}{m_{\text{col}}}\right) \right. \nonumber \\
    &\hspace{14.57em}-
    \frac{1}{2}C_q\left(x_2; x,\frac{M_{\text{UV}}}{m_{\text{col}}}\right) -
    \frac{1}{2}C_q\left(x_2'; x,\frac{M_{\text{UV}}}{m_{\text{col}}}\right) \nonumber \\
    &\hspace{14.57em}-
    \left.\! \frac{1}{2}C_q\left(x_3; x,\frac{M_{\text{UV}}}{m_{\text{col}}}\right) -
    \frac{1}{2}C_q\left(x_3'; x,\frac{M_{\text{UV}}}{m_{\text{col}}}\right)\right]\ .
  \label{eq:renormfactors}
\end{align}
Here the $x_i$ should be read as shorthands for the associated
expressions in terms of $\xi$ and $\eta$ as defined in
\eqref{eq:xietadef}, i.e.\ $x_1=\xi\eta$, $x_2=\eta(1-\xi)$, and
$x_3=1-\eta$. For clarity from now on we shall write the arguments
of the three-quark wavefunction in the following order:
$\Psi_{\text{qqq}}(x_1,x_2,\vec k_1, \vec k_2)$. It is understood that the
third LC momentum fraction and transverse momentum is determined by
conservation of momentum.

Tracing over the quark transverse momenta $\vec{k}_i$ then gives e.g.\
\begin{align}
  \rho^{(11)}_{x_1 x_2,x_1'x_2'} &= -\frac{g^2C_F}{(4\pi)^2}\,
  \log\frac{M_{\text{UV}}^2}{m_{\text{col}}^2}\,\,
\frac{1}{4}
\int [\dd ^2k_i] \int_x^1 \frac{\dd x_g}{x_g} \
\Theta(x_1-x_g)\,
\left[ 1 + \left(1 - z\right)^2\right]
\frac{\Psi^*_{\text{qqq}}(x_1',x_2',\vec{k}_1,\vec{k}_2)}{2\sqrt{x_1' x_2' x_3'}} \,
\frac{\Psi_{\text{qqq}}(x_1,x_2,\vec{k}_1,\vec{k}_2)}
     {2\sqrt{x_1 x_2 x_3}} \\
&=
 -\frac{g^2C_F}{(4\pi)^2}\,
 \log\frac{M_{\text{UV}}^2}{m_{\text{col}}^2}\,
  \int_x^1 \frac{\dd x_g}{x_g} \
\Theta(x_1-x_g)\,
\left[ 1 + \left(1 - z\right)^2\right]
\,\rho^\mathrm{qqq}_{x_1 x_2,x_1'x_2'}~.
\label{eq:rhoqqqg11-x1x2}
\end{align}
$\rho^\mathrm{qqq}_{x_1 x_2,x_1'x_2'}$ is the non-perturbative
reduced density matrix of the three quark state before
gluon emission, see sec.~\ref{sec:dPDF_qqq}.
The trace of this density matrix is computed by setting $x_1=x_1'$,
$x_2 = x_2'$ and integrating with the measure $\dd x_1 \dd x_2$, as
already mentioned in eq.~\eqref{eq:tr_x1-x2}.  Changing to
unconstrained $\xi$, $\eta$ variables,
\begin{align}
\rho^{(11)}_{\xi\eta,\xi'\eta'} &= -\frac{g^2C_F}{(4\pi)^2}\,
  \log\frac{M_{\text{UV}}^2}{m_{\text{col}}^2}\,\,
\frac{1}{4}
\int [\dd ^2k_i] \int_x^1 \frac{\dd x_g}{x_g} \, \Theta(\xi\eta-x_g)
\left[ 1 + \left(1 - z\right)^2\right] \nonumber\\
& ~~~~~~~~~~~~~
\frac{\Psi^*_{\text{qqq}}(\xi'\eta',\eta'(1-\xi'),\vec{k}_1,\vec{k}_2)}
{\sqrt{4\eta'(1-\eta')\xi'(1-\xi')}}\,
\frac{\Psi_{\text{qqq}}(\xi\eta,\eta(1-\xi),\vec{k}_1,\vec{k}_2)}
{\sqrt{4\eta(1-\eta)\xi(1-\xi)}}~.
\label{eq:rhoqqqg11}
\end{align}
The trace of this density matrix is computed by setting $\xi=\xi'$,
$\eta =\eta'$ and integrating with the measure
$\dd\xi\dd\eta$,
cf.\ eqs.~(\ref{eq:measure_x1-x2-->xi-eta},\ref{eq:xietatrace}).
Adding this contribution to $\rho_{\xi\eta,\xi'\eta'}$ on the
r.h.s.\ of eq.~(\ref{eq:rho-red_xi-eta}) accounts for the ${\cal
  O}(\alpha_s\log{M_{\text{UV}}^2}/{m_{\text{col}}^2})$ virtual correction.  \\

Now we consider the real emission correction between quarks $i$ and
$i'$. The four particle density matrix for the $1$ to $1'$ gluon
exchange (summed over gluon polarizations) is
\begin{align}
  \rho^{(11')}_{\text{qqqg}} &=
  \frac{g^2}{3}\sum_{m,m'}\epsilon_{m'n_2'n_3'}\,(t^{a'})_{m'n_1'}\,
  \epsilon_{mn_2n_3}\,(t^a)_{n_1m}\,
  \Theta(x_1-x_g)\,\Theta(x_1'-x_g') \nonumber \\
&\hspace{10em}\times
\frac{\vec{n}\cdot \vec{n}'}{(n^2+\Delta^2)(n'^2+\Delta'^2)}\,
  (2-z-z'+zz')\,
  \frac{\Psi^*_{\text{qqq}}(x'_1,x'_2,\vec{k}'_1,\vec{k}'_2)}
  {2\sqrt{x_1' x_2' x_3'}}\,
  \frac{\Psi_{\text{qqq}}(x_1,x_2,\vec{k}_1,\vec{k}_2)}
  {2\sqrt{x_1 x_2 x_3}}
\end{align}
with $z = x_g/x_1$, $\Delta$ defined as before, $\vec{n} =
\vec{k}_g-z\vec{k}_1$, and the primed variables defined
analogously. Here we have used the Hilbert space basis
introduced in eq.~\eqref{eq:basis_|alpha>}.

We need $x_1$ to refer to the daughter quark LC momentum
when we project onto basis states, so we shift $x_1 \mapsto x_1 +
x_g$. Doing this, then tracing over the gluon color and momentum as
well as over the parent quark transverse momenta, yields
\begin{align}
  \rho^{(11')}_{x_1 x_2, x_1' x_2'} &= {2g^2C_F}\,\frac{1}{4}
\int [\dd ^2k_i] 
\int_x^1 \frac{\dd x_g}{x_g}\,\Theta(1-x_1-x_2-x_g)\,\Theta(1-x_1'-x_2'-x_g)
\intddkg \nonumber \\
  &\hspace{5em}\times
  \frac{\vec{n}\cdot \vec{n}'}{(n^2+\Delta^2)(n'^2+\Delta'^2)}\,(2-z-z'+zz')\,
  \frac{\Psi^*_{\text{qqq}}(x_1'+x_g,x'_2,\vec{k}_1,\vec{k}_2)}
  {2\sqrt{(x_1'+x_g) x_2' x_3'}}\,
  \frac{\Psi_{\text{qqq}}(x_1+x_g,x_2,\vec{k}_1,\vec{k}_2)}
  {2\sqrt{(x_1 +x_g) x_2 x_3}}~,
\end{align}
where now $z = x_g/(x_1+x_g)$, $z' = x_g/(x_1'+x_g)$ due to the
shifted arguments, and $x_3=1-x_1-x_2-x_g$ (and similar for
$x_3'$).

We then introduce the c.o.m.\ variables $\xi$ and $\eta$ in the same
way as in \eqref{eq:xietadef} noting that now $x_3 = 1-\eta-x_g$ and
$\eta < 1-x_g$.
Furthermore,
\begin{align}
  \frac{ \dd x_1\, \dd x_2\, \dd x_3 }{4(x_1+x_g)x_2x_3}\,
  \delta(1 - x_1 - x_2 - x_3 - x_g) =
  \frac{\dd\xi}{2\xi(1-\xi)}
  \frac{\dd\eta}{2\eta(1-\eta)}
  \frac{ \Theta(1 - \eta - x_g) }{\left(1+\frac{x_g}{\xi\eta}\right)\left(1-\frac{x_g}{1-\eta}\right)}
  \label{eq:measure_x1-x2_xi-eta_xg}
\end{align}
so that
\begin{align}
  \rho^{(11')}_{\xi\eta, \xi' \eta'} &= {2g^2C_F}\,\frac{1}{4}
\int [\dd ^2k_i] 
\int_x^1 \frac{\dd x_g}{x_g}
\frac{ \Theta(1 - \eta - x_g) }{\sqrt{\left(1+\frac{x_g}{\xi\eta}\right)\left(1-\frac{x_g}{1-\eta}\right)}}\,
\frac{ \Theta(1 - \eta' - x_g) }{\sqrt{\left(1+\frac{x_g}{\xi'\eta'}\right)\left(1-\frac{x_g}{1-\eta'}\right)}}
\intddkg \frac{\vec{n}\cdot \vec{n}'}{(n^2+\Delta^2)(n'^2+\Delta'^2)}\nonumber \\
&\hspace{8.5em}\times (2-z-z'+zz')\,
\frac{\Psi^*_{\text{qqq}}(\xi'\eta'+x_g,\eta'(1-\xi'),\vec{k}_1,\vec{k}_2)}{\sqrt{4\eta'(1-\eta')\xi'(1-\xi')}}\,
\frac{\Psi_{\text{qqq}}(\xi\eta+x_g,\eta(1-\xi),\vec{k}_1,\vec{k}_2)}
{\sqrt{4\eta(1-\eta)\xi(1-\xi)}}
\end{align}
with $z = x_g/(\xi\eta+x_g)$, $z' = x_g/(\xi' \eta'+x_g)$.
We verify that for $(\xi,\eta)\in (0,1)^2$ the arguments of the
three-quark wavefunctions do not take unphysical values: $x_g < 1-\eta$
implies $\xi\eta+x_g < (\xi-1)\eta+1 < 1$.

The previous expression is logarithmically UV divergent and we
again introduce a UV regulator as
\begin{align}
 & -{g^2C_F}\, 
\frac{1}{4}
\int [\dd ^2k_i] \int_x^1 \frac{\dd x_g}{x_g}
\frac{ \Theta(1 - \eta - x_g) }{\sqrt{\left(1+\frac{x_g}{\xi\eta}\right)\left(1-\frac{x_g}{1-\eta}\right)}}\,
\frac{ \Theta(1 - \eta' - x_g) }{\sqrt{\left(1+\frac{x_g}{\xi'\eta'}\right)\left(1-\frac{x_g}{1-\eta'}\right)}}
\int \frac{\dd^2 k_g}{16\pi^3}
\, (2-z-z'+zz')
\nonumber \\
&\hspace{6em}\times \left[
\frac{1}{n^2+\Lambda^2} +
\frac{1}{n'^2+\Lambda^{\prime 2}}
\right]
\frac{\Psi^*_{\text{qqq}}(\xi'\eta'+x_g,\eta'(1-\xi'),\vec{k}_1,\vec{k}_2)}{\sqrt{4\eta'(1-\eta')\xi'(1-\xi')}}\,
\frac{\Psi_{\text{qqq}}(\xi\eta+x_g,\eta(1-\xi),\vec{k}_1,\vec{k}_2)}
     {\sqrt{4\eta(1-\eta)\xi(1-\xi)}}
~.
\label{eq:rhoqqqg11real}
\end{align}
All together we have
\begin{align}
\rho^{(11')}_{\xi\eta,\xi'\eta'} &= {2g^2C_F} \frac{1}{4}
\int [\dd ^2k_i] \int_x^1 \frac{\dd x_g}{x_g} \,
\frac{ \Theta(1 - \eta - x_g) }{\sqrt{\left(1+\frac{x_g}{\xi\eta}\right)\left(1-\frac{x_g}{1-\eta}\right)}}\,
\frac{ \Theta(1 - \eta' - x_g) }{\sqrt{\left(1+\frac{x_g}{\xi'\eta'}\right)\left(1-\frac{x_g}{1-\eta'}\right)}} (2-z-z'+zz')\intddkg\nonumber \\
&\hspace{1em}\times
\left[ \frac{\vec{n}\cdot \vec{n}'}{(n^2+\Delta^2)(n'^2+\Delta'^2)} - \frac{1}{2}\frac{1}{n^2 + \Lambda^2} - \frac{1}{2}\frac{1}{n'^2 + \Lambda'^2} \right]
\frac{\Psi^*_{\text{qqq}}(\xi'\eta'+x_g,\eta'(1-\xi'),\vec{k}_1,\vec{k}_2)}{\sqrt{4\eta'(1-\eta')\xi'(1-\xi')}}\,
\frac{\Psi_{\text{qqq}}(\xi\eta+x_g,\eta(1-\xi),\vec{k}_1,\vec{k}_2)}
     {\sqrt{4\eta(1-\eta)\xi(1-\xi)}}
     ~.
\label{eq:rhoqqqg11'}
\end{align}
At this point we pause to verify not only that the trace of the density matrix
is independent of the collinear and UV ($m_{\text{col}}$,
$M_{\text{UV}}$) and soft ($x$) cutoffs but, in fact, that the ${\cal
  O}(\alpha_s)$ correction cancels altogether so as to preserve the
normalization $\tr\, \rho=1$.

To compute the trace of~\eqref{eq:rhoqqqg11'} we set $\xi'=\xi$,
$\eta'=\eta$, and integrate over ${\dd\xi}{\dd\eta}$. We can then
proceed in two ways. First, we can use
eq.~\eqref{eq:measure_x1-x2_xi-eta_xg} in reverse thereby also
replacing $\xi\eta\to x_1$, $\eta(1-\xi)\to x_2$, implying
$1-\eta-x_g\to x_3$.  We then undo the shift of $x_1$ by letting $x_1
\to x_1-x_g$ which also introduces a $\Theta(x_1-x_g)$ function. The
resulting expression cancels (at leading logarithmic accuracy, up to
power corrections) the trace of $\rho^{(11)}$ from
eq.~\eqref{eq:rhoqqqg11-x1x2} plus the same contribution from the
trace of $\rho^{(1'1')}$.

Alternatively, we can also verify the cancellation of the integrals
for $\tr \rho^{(11')} + \tr \rho^{(11)} + \tr \rho^{(1'1')}$ in the
$\xi, \eta$ variables. In $\tr \rho^{(11')}$ we perform the
transformation
\begin{equation}
  \tilde\eta = \eta + x_g~~,~~ \tilde\xi = \frac{\xi\eta+x_g}{\eta+x_g}
\end{equation}
so that $1-\eta-x_g=1-\tilde\eta$,
$\eta(1-\xi)=\tilde\eta(1-\tilde\xi)$, and $\xi\eta+x_g=\tilde\xi
\tilde\eta$ (similar for the primed variables). This gives rise to a
factor $\Theta(\tilde\xi\tilde\eta-x_g)$, which allows us to extend
the range of both $\tilde\xi$ and $\tilde\eta$ to $[0,1]$, and
restores $z=x_g/\tilde\xi\tilde\eta$ as well as
$\Psi_{\text{qqq}}(\xi\eta+x_g,\eta(1-\xi),\vec{k}_1,\vec{k}_2) =
\Psi_{\text{qqq}}(\tilde\xi\tilde\eta,\tilde\eta(1-\tilde\xi),\vec{k}_1,\vec{k}_2)$. Furthermore,
\begin{equation}
  \frac{\dd\eta\, \dd\xi}{2\xi(1-\xi)\, 2\eta(1-\eta)}
  \frac{1}{\left(1+\frac{x_g}{\xi\eta}\right)\left(1-\frac{x_g}{1-\eta}\right)} =
  \frac{\dd\tilde\eta\, \dd\tilde\xi}{2\tilde\xi(1-\tilde\xi)\,
    2\tilde\eta(1-\tilde\eta)}
\end{equation}
Thus, we again obtain that the trace of eq.~\eqref{eq:rhoqqqg11'}
cancels against $\tr \rho^{(11)}$ (plus the same
contribution from $\tr \rho^{(1'1')}$).
\\~~\\

We now continue our computation with the contribution from the $2$ to
$2'$ gluon exchange. Here we shift $x_2\to x_2+x_g$ so that $x_2$
now corresponds to the LC momentum fraction of daughter quark 2.
The measure changes as
\begin{align}
\frac{ \dd x_1\, \dd x_2\, \dd x_3 }{4x_1(x_2+x_g)x_3}\,\delta(1 - x_1 - x_2 - x_3 - x_g)
  = \frac{\dd\xi}{2\xi(1-\xi)}
  \frac{\dd\eta}{2\eta(1-\eta)}\,
  \frac{ \Theta(1 - \eta - x_g) }{\left(1+\frac{x_g}{(1-\xi)\eta}\right)\left(1-\frac{x_g}{1-\eta}\right)}~.
\end{align}
Then
\begin{align}
\rho^{(22')}_{\xi\eta,\xi'\eta'} &= {2g^2C_F} \frac{1}{4}
\int [\dd ^2k_i] \int_x^1 \frac{dx_g}{x_g} \intddkg\,
\frac{ \Theta(1 - \eta - x_g) }{\sqrt{\left(1+\frac{x_g}{(1-\xi)\eta}\right)\left(1-\frac{x_g}{1-\eta}\right)}}\,
\frac{ \Theta(1 - \eta' - x_g) }{\sqrt{\left(1+\frac{x_g}{(1-\xi')\eta'}\right)\left(1-\frac{x_g}{1-\eta'}\right)}} (2-z-z'+zz')\nonumber \\
&\hspace{1em}\times  \left[ \frac{\vec{n}\cdot \vec{n}'}{(n^2+\Delta^2)(n'^2+\Delta'^2)} - \frac{1}{2}\frac{1}{n^2 + \Lambda^2} - \frac{1}{2}\frac{1}{n'^2 + \Lambda'^2} \right]\,
\frac{\Psi^*_{\text{qqq}}(\xi'\eta',\eta'(1-\xi')+x_g,\vec{k}_1,\vec{k}_2)}
{\sqrt{4\eta'(1-\eta')\xi'(1-\xi')}}\,
\frac{\Psi_{\text{qqq}}(\xi\eta,\eta(1-\xi)+x_g,\vec{k}_1,\vec{k}_2)}
{\sqrt{4\eta(1-\eta)\xi(1-\xi)}}
\ ,
\label{eq:rhoqqqg22'}
\end{align}
where now $z = x_g/(\eta(1-\xi)+x_g)$, $z' = x_g/(\eta'(1-\xi')+x_g)$,
$\vec{n} = \vec{k}_g - z\vec{k}_2$, and $\Delta$, $\Lambda$ and the
corresponding primed variables defined as before. For this
contribution the transformation
\begin{equation}
  \tilde\eta = \eta+x_g~~,~~ \tilde\xi = \frac{\xi\eta}{\eta+x_g}~,
\end{equation}
which corresponds to $\tilde\xi\tilde\eta=\xi\eta$,
$\tilde\eta(1-\tilde\xi)=\eta(1-\xi)+x_g$, is useful for
checking the cancellation of the perturbative correction to
the trace.\\

Finally, for the gluon exchange from $3$ to $3'$, no shift from parent
to daughter quark LC momentum is needed as we are interested in the
reduced density matrix over quarks $1$ and $2$. The density matrix for
this diagram is
\begin{align}
\rho^{(33')}_{\xi\eta,\xi'\eta'} &= {2g^2C_F} \frac{1}{4}
\int [\dd ^2k_i] \int_x^1 \frac{dx_g}{x_g} \intddkg\, \Theta(1 - \eta - x_g) \, \Theta(1 - \eta' - x_g)  \, (2-z-z'+zz')\nonumber \\
&\hspace{1em}\times  \left[ \frac{\vec{n}\cdot \vec{n}'}{(n^2+\Delta^2)(n'^2+\Delta'^2)} - \frac{1}{2}\frac{1}{n^2 + \Lambda^2} - \frac{1}{2}\frac{1}{n'^2 + \Lambda'^2} \right]\,
\frac{\Psi^*_{\text{qqq}}(\xi'\eta',\eta'(1-\xi'),\vec{k}_1,\vec{k}_2)}
{\sqrt{4\eta'(1-\eta')\xi'(1-\xi')}}\,
\frac{\Psi_{\text{qqq}}(\xi\eta,\eta(1-\xi),\vec{k}_1,\vec{k}_2)}
{\sqrt{4\eta(1-\eta)\xi(1-\xi)}}
\ ,
\label{eq:rhoqqqg33'}
\end{align}
with $z=x_g/(1-\eta)$, $z'=x_g/(1-\eta')$. Here the cancellation
of the trace against the last line in~\eqref{eq:renormfactors}
is seen immediately.\\

In summary, single step evolution of the three-quark density matrix
from the scale $m^2_{\text{col}}=Q_0^2$ to the higher scale
$M^2_{\text{UV}}=Q^2$ is performed by replacing
$\Psi^*_{\text{qqq}}(\xi'\eta',\eta'(1-\xi'),\vec{k}_1,\vec{k}_2)\,
\Psi_{\text{qqq}}(\xi\eta,\eta(1-\xi),\vec{k}_1,\vec{k}_2)$
in~\eqref{eq:rho-red_xi-eta} by the sum of virtual
corrections~\eqref{eq:renormfactors} followed by adding the density
matrices~(\ref{eq:rhoqqqg11'}, \ref{eq:rhoqqqg22'},
\ref{eq:rhoqqqg33'}) for real emissions.  Our result applies in the
leading $\log M_{\text{UV}}/m_{\text{col}} \gg1$ approximation when
the coupling $\alpha_s$ is sufficiently small, and perhaps the cutoff
$x$ on the LC momentum fraction of the gluon is sufficiently large, so
that the ${\cal O}(\alpha_s)$ contribution to the wavefunction
renormalization factor~\eqref{eq:cqdef} is small.

\subsection{Cancellation of Soft Divergence}

In the previous section, we had to introduce a cutoff $x$ on the
LC momentum of the gluon to handle the soft divergence. Here we 
show that this divergence cancels at leading logarithmic accuracy
in the sum of real emissions and virtual corrections.
Hence, the integration over $x_g$ could in principle extend to 0.

The expressions from the previous section simplify greatly when $x_g$
is much less than the typical momentum fraction of a quark,
$\exv{x_{\text{q}}}$. We then have $z, z' \ll 1$.
The integral over the gluon transverse momentum $\vec{k}_g$
for the virtual correction
in eq.~\eqref{eq:rhoqqqg11} becomes
\begin{align}
  \intddkg
  \left( \frac{1}{k_g^2 + \Delta^2} -
  \frac{1}{k_g^2+\Lambda^2}\right)
= \frac{1}{(4\pi)^2} \log \frac{M_{\text{UV}}^2}{m_{\text{col}}^2}
\label{eq:rhoqqqg11smallx}\ .
\end{align}
For the real exchanges, the softness of the gluon implies that the
daughter quark momentum is the same as that of the parent quark. Since
$z,z' \ll 1$, we have $\vec{n} = \vec{n}' = \vec{k}_g$ and
the integral over the gluon transverse momentum
in $\rho^{(11')}$ from eq.~(\ref{eq:rhoqqqg11'}) becomes
\begin{align}
2 \intddkg
\left[ \frac{k_g^2}{(k_g^2+\Delta^2)(k_g^2+\Delta'^2)} -
  \frac{1}{2}\frac{1}{k_g^2 + \Lambda^2} -
  \frac{1}{2}\frac{1}{k_g^2 + \Lambda'^2} \right]
\ . 
\end{align}
Consider now the sum of the gluon exchange between quarks 1 and $1'$
plus the corresponding virtual corrections
in the small $x_g$ limit:
\begin{align}
\intddkg
\left[2\left(\frac{k_g^2}{(k_g^2+\Delta^2)(k_g^2+\Delta'^2)} -
  \frac{1}{2}\frac{1}{k_g^2 + \Lambda^2} -
  \frac{1}{2}\frac{1}{k_g^2 + \Lambda'^2}\right)
  -\left( \frac{1}{k_g^2 + \Delta^2} -
  \frac{1}{k_g^2+\Lambda^2} \right)  -
  \left( \frac{1}{k_g^2 + \Delta'^2} -
  \frac{1}{k_g^2+\Lambda'^2} \right) \right]\ .
\label{eq:gluonsum1}
\end{align}
Note that
\begin{equation}
\frac{2k_g^2}{(k_g^2+\Delta^2)(k_g^2+\Delta'^2)} = \frac{1}{k_g^2+\Delta^2} + \frac{1}{k_g^2+\Delta'^2} + \frac{\Delta^2 + \Delta'^2}{2(k_g^2+\Delta^2)(k_g^2+\Delta'^2)}~. \label{eq:kgfractrick}
\end{equation}
The last term does not contribute to the leading-log approximation and
can be dropped. This results in a cancellation of the bracketed terms
in \eqref{eq:gluonsum1}, hence we can take $x \to 0$ with no overall
soft divergence when considering the sum of these diagrams. No trace
over $\xi$ or $\eta$ has been performed here: this cancellation occurs
for the entire density matrix, not just the diagonal elements. The
calculation is identical for quarks 2 and $2'$, and 3 and $3'$.

\subsection{DGLAP evolution of the diagonal of the density matrix}
\label{sec:rho_diagonal-DGLAP}

In this section we obtain expressions for the derivative with respect 
to\ $Q^2=M_{\text{UV}}^2$ of the diagonal elements of the density
matrix, i.e.\ the double quark PDF. These can be expressed in terms
of convolutions of splitting functions and dPDFs.

We have from eq.~\eqref{eq:rhoqqqg11-x1x2}, using relations~(\ref{eq:proton-dm},\ref{eq:reduced-rho_x1-x2}),
\begin{equation}
  Q^2\frac{\partial}{\partial Q^2} \rho^{(11)}_{x_1 x_2,x_1x_2} =
  Q^2\frac{\partial}{\partial Q^2} \rho^{(1'1')}_{x_1 x_2,x_1x_2} =
  - \frac{\alpha_s}{4\pi}\int\limits^1_{x/x_1}\dd z \,
  P_{g\leftarrow q}(z)\, \rho^\mathrm{qqq}_{x_1 x_2,x_1x_2}~.
\end{equation}
$\rho^\mathrm{qqq}_{x_1 x_2,x_1'x_2'}$ denotes the non-perturbative
reduced density matrix obtained from the three quark LCwf before
gluon emission, see sec.~\ref{sec:dPDF_qqq}.
Similarly,
\begin{align}
  Q^2\frac{\partial}{\partial Q^2} \rho^{(22)}_{x_1 x_2,x_1x_2} & =
  Q^2\frac{\partial}{\partial Q^2} \rho^{(2'2')}_{x_1 x_2,x_1x_2} =
  - \frac{\alpha_s}{4\pi}\int\limits^1_{x/x_2}\dd z \,
  P_{g\leftarrow q}(z)\, \rho^\mathrm{qqq}_{x_1 x_2,x_1x_2}~,\\
  Q^2\frac{\partial}{\partial Q^2} \rho^{(33)}_{x_1 x_2,x_1x_2} & =
  Q^2\frac{\partial}{\partial Q^2} \rho^{(3'3')}_{x_1 x_2,x_1x_2} =
  - \frac{\alpha_s}{4\pi}\int\limits^1_{\frac{x}{1-x_1-x_2}}\dd z \,
  P_{g\leftarrow q}(z)\, \rho^\mathrm{qqq}_{x_1 x_2,x_1x_2}~.
\end{align}

Continuing with the real emission corrections,
\begin{align}
  Q^2\frac{\partial}{\partial Q^2} \rho^{(11')}_{x_1 x_2,x_1x_2} &=
  \frac{2\alpha_s}{4\pi} \int \limits_{x_1/(1-x_2)}^{x_1/(x_1+x)}
  \frac{\dd z}{z}\, P_{g\leftarrow q}(1-z)\,
  \rho^\mathrm{qqq}_{\frac{x_1}{z} x_2,\frac{x_1}{z}x_2}~,
  \label{eq:D_t-rho11'}\\
Q^2\frac{\partial}{\partial Q^2} \rho^{(22')}_{x_1 x_2,x_1x_2} &=
  \frac{2\alpha_s}{4\pi} \int \limits_{x_2/(1-x_1)}^{x_2/(x_2+x)}
  \frac{\dd z}{z}\, P_{g\leftarrow q}(1-z)\,
  \rho^\mathrm{qqq}_{x_1\frac{x_2}{z},x_1\frac{x_2}{z}}~,
  \label{eq:D_t-rho22'}\\
  Q^2\frac{\partial}{\partial Q^2} \rho^{(33')}_{x_1 x_2,x_1x_2} &=
  \frac{2\alpha_s}{4\pi}\int\limits^1_{\frac{x}{1-x_1-x_2}}\dd z \,
  P_{g\leftarrow q}(z)\, \rho^\mathrm{qqq}_{x_1 x_2,x_1x_2}~.
\end{align}
The last term simply cancels the corresponding virtual corrections.
\\

The expressions above reproduce the standard equations for DGLAP
evolution of double quark PDFs\footnote{For a single quark flavor, and
with the initial gluon PDF set to 0.} with the identification
$f_{qq}(x_1,x_2)= \rho_{x_1 x_2,x_1x_2}$, which agrees with
eq.~\eqref{eq:dPDF-red_rho}. The
standard form of the dPDF DGLAP equations is
\begin{equation}
  \partial_t f_{qq}(x_1,x_2) = \int\limits_0^{1-x_2}\dd u \, K(x_1,u)\,
  f_{qq}(u,x_2) +
   \int\limits_0^{1-x_1}\dd u \, K(x_2,u)\, f_{qq}(x_1,u)~,
\end{equation}
where $t=\log Q^2$. For example, the contribution involving the real emission
kernel
\begin{equation}
K_R(x,u) = \frac{\alpha_s}{2\pi} \frac{1}{u} P_{qq}(x/u)\, \Theta(u-x)
\end{equation}
corresponds exactly to the sum of~(\ref{eq:D_t-rho11'},
\ref{eq:D_t-rho22'}). To see this, use the relation $P_{gq}(1-z) =
P_{qq}(z)$ and the transformation $u=x_1/z$ or $u=x_2/z$, and finally
send the cutoff $x\to 0$.  The virtual corrections serve to ensure
conservation of probability.

Hence, the diagonal of the density matrix evolves according to dPDF DGLAP
via a convolution with splitting functions,
and independently of the off-diagonal elements.

\subsection{Numerical Results}

In this section we present first qualitative results on the evolution
of quantum correlations in the dPDF to higher scales by evaluating the
correction to the density matrix due to the emission of one collinear
gluon. The full evolution equation would sum multiple parton splitting
steps, a more involved task which we defer to future work.

To keep the perturbative correction small we employ an unrealistically
small coupling, $\alpha_s=0.1$, and a fairly large cutoff $x=0.1$ on
the LC momentum fraction of the gluon.  The initial scale $Q_0$
corresponding to the quark mass regulator $m_{\text{col}}$ is set to
the quark mass $m_q = 0.26\ \text{GeV}$ encoded in the
non-perturbative three-quark BS wavefunction presented at the end of
sec.~\ref{sec:qqq-state}.  Also, we choose $Q=M_{\text{UV}}=10
m_{\text{col}}$.

The computation of the evolved density matrix is numerically very
challenging. It requires the evaluation of a seven dimensional
integral, e.g.\ eq.~\eqref{eq:rhoqqqg11'}, at each of $N^2 \times N^2
/ 2$ entries of the $(\xi,\eta)-(\xi',\eta')$ symmetric density
matrix, where $N$ is the number of bins per variable.  We have used
the vegas+ v6.2 algorithm~\cite{Lepage:2020tgj,vegas-6.2} to
estimate the density matrix with collinear gluon emission corrections.
The evolved dPDF, as the diagonal of the density matrix,
only requires $N^2$ integrations, so we have computed it with $N=100$ bins.
On the other hand, studying the effects of the PEN transformation 
requires the entire density matrix, limiting us to $N=20$ bins.
For this reason we refrain from showing the dPDF to sPDF ratio $C(x_1,x_2)$
but focus instead on the dPDF itself.

\begin{figure}[htb]
  \includegraphics[width=0.48\textwidth]{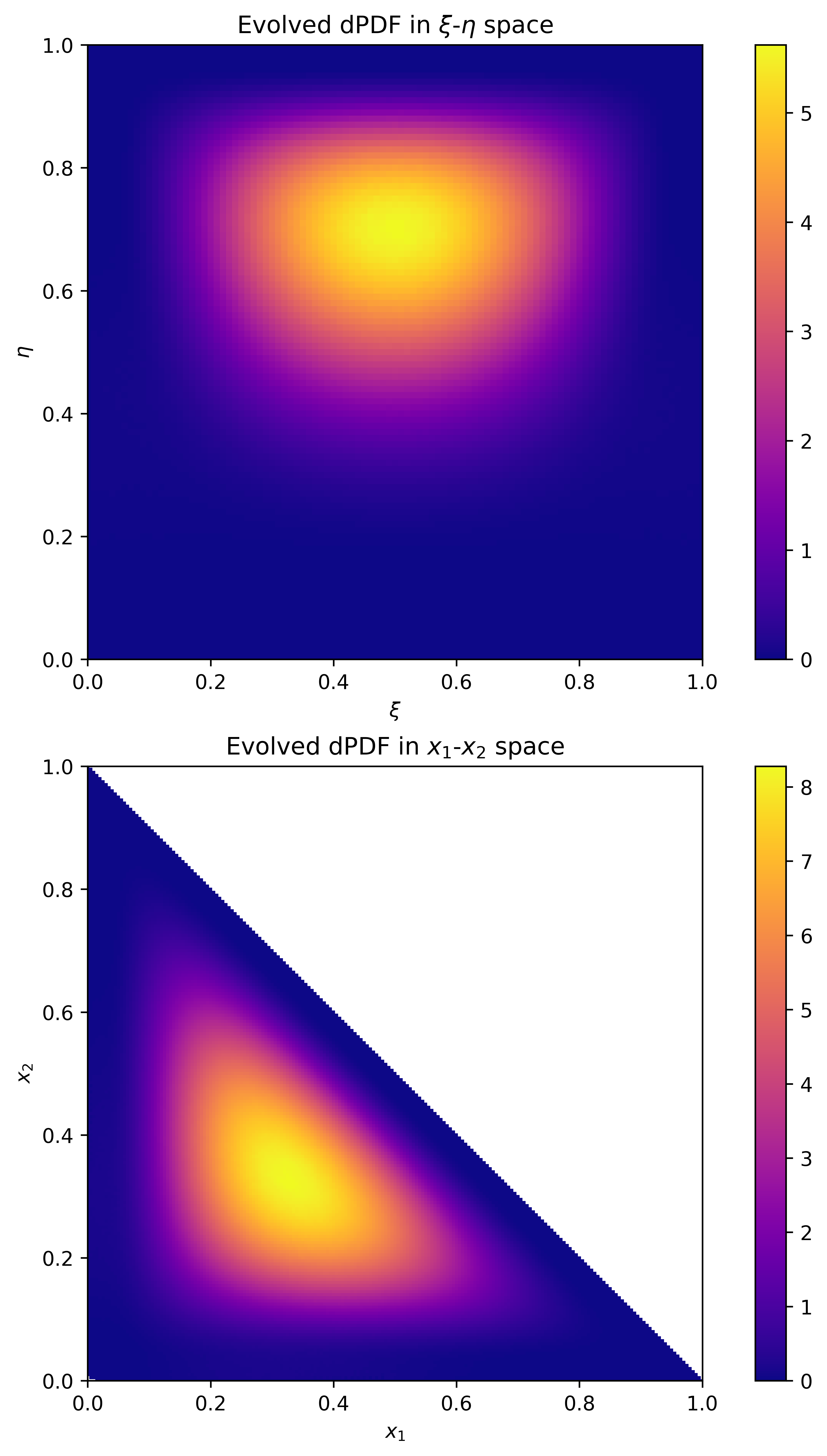}
  \includegraphics[width=0.48\textwidth]{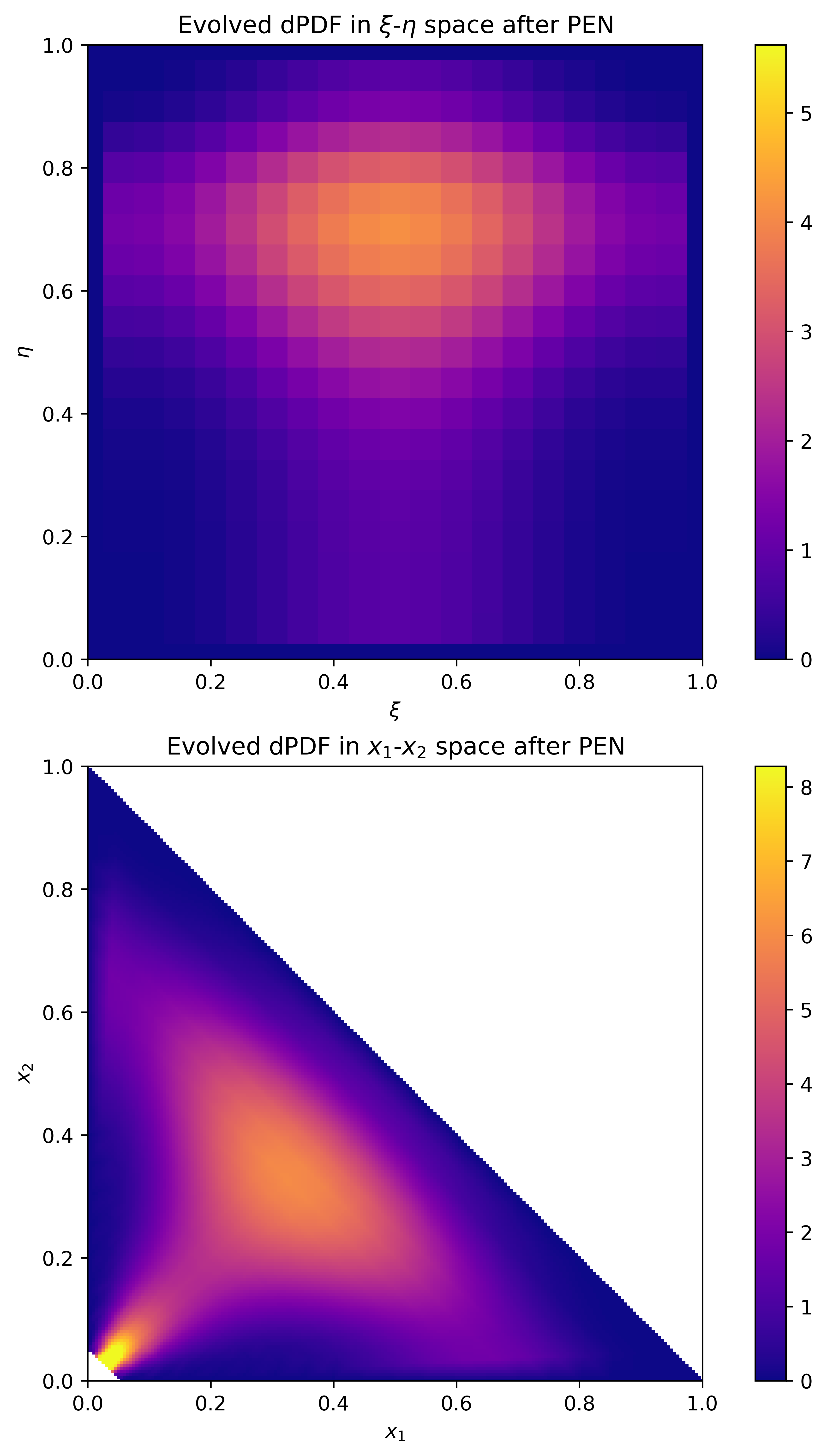}
  \vspace*{-.3cm}
  \caption{Contour plots of the dPDF at $Q^2/Q^2_0 = 100$ with $\alpha_s = 0.1$, $x=0.1$, and $Q_0 = m_q$ as a function
    of $\eta=x_1+x_2$ and $\xi = x_1/(x_1+x_2)$ (top), and $x_1$ and $x_2$ (bottom). Left: the dPDF
    obtained here from the Brodsky-Schlumpf model LCwf with
    corrections due to collinear gluon emission. Right: same, after the
    removal of quantum correlations via the PEN transformation.}
  \label{fig:dPDF-slices-nlo}
\end{figure}
Our numerical results are shown in fig.~\ref{fig:dPDF-slices-nlo}.
The maximum of the dPDF occurs at $\xi=1/2$, of course, and near
$\eta=2/3$.  The strength of this peak is reduced by the PEN
transformation.  On the other hand, it enhances the dPDF for
asymmetric momentum fractions (small or large $\xi$) when either one
of them is large ($\eta=2/3 \pm 0.15$). PEN also enhances the dPDF for
similar ($\xi=1/2 \pm 0.15$) but fairly small ($\eta\lesssim 0.3$)
momentum fractions.  Put the other way: entanglement correlations
associated with the negativity tend to suppress the dPDF for small
(and similar) as well as for large but strongly asymmetric momentum
fractions, but instead strengthen the peak near $x_1=x_2$ and
$x_1+x_2=2/3$.  Thus, after evolution to higher $Q^2$ the strongest
effects from quantum correlations appear in a different regime than at
the initial scale $Q_0^2$; we recall from
sec.~\ref{sec:dPDF-results-LO} that for the non-perturbative
constituent quark LCwf the strongest effects occurred for rather
asymmetric momentum fractions $x_1$ and $x_2$, and well below the
kinematic boundary $x_1+x_2=1$.

\section{Summary}

The density matrix $\rho_{x_1 x_2, x_1' x_2'}$ describing the LC momentum
distributions of two quarks in the proton is obtained by tracing the
pure proton state $\ket{P}\, \bra{P}$ over all unobserved degrees of freedom.
This, in general, results in a mixed state with non-zero von~Neumann
entropy. The purpose of this paper was to apply methods from
Quantum Information Theory, specifically the negativity measure, to
investigate quantum entanglement correlations of double quark momentum
fractions.

The diagonal $f_{qq}(x_1,x_2)=\rho_{x_1 x_2, x_1 x_2}$ of the density
matrix corresponds to the double quark parton distribution (dPDF). In
general the dPDF differs from the product of two single quark PDFs due
to correlations. Classical correlations can be encoded in
probabilistic mix of product states, i.e.\ a separable
state~\eqref{eq:rho_separable}; for such a state the ``entanglement
negativity'' measure of Quantum Information Theory is zero.  On the
other hand, a non-zero negativity implies the presence of
non-separable quantum correlations.

In sec.~\ref{sec:dPDF_qqq} we have computed the density matrix over
the two-quark LC momentum fractions in an effective LC constituent
quark model of the proton intended for not too small $x_i \sim 0.1$
or greater, and a low resolution scale $Q_0^2$. (For an
improved description of the proton at low scale one should account
for non-perturbative contributions from higher Fock states such as
$|qqqg\rangle$ and $|qqqq\bar{q}\rangle$ in the future.) Upon purging the
density matrix of its entanglement negativity and the associated
quantum correlations, we observe strong modifications of the dPDF in
some range of momentum fractions, specifically for asymmetric $x_1,
x_2$ far below the bound $x_1+ x_2=1$. In fact, in this regime we find
that purging the entanglement negativity turns a suppression of the
quantum correlated dPDF (relative to a product of two single quark
PDFs) into a strong enhancement.  On the other hand, for large $x_1+
x_2$ or $x_1\simeq x_2$ the effect of such quantum correlations in
this model is found to be small whereas classical correlations
enforce $f_{qq}(x_1,x_2)/f_q(x_1) f_q(x_2)\to 0$ as $x_1+ x_2\to 1$.

In sec.~\ref{sec:rho_evol} we derived the corrections to $\rho_{x_1
  x_2, x_1' x_2'}$ due to the emission of one collinear gluon. This
corresponds to one step of QCD scale evolution to $Q^2 > Q_0^2$ of the
entire density matrix, assuming $\alpha_s \log Q^2/Q_0^2 \ll 1$; on
the diagonal we recover the dPDF DGLAP evolution in terms of
convolutions of splitting functions and dPDFs. From the density matrix
at the new scale $Q^2$ we obtain first qualitative predictions for the
QCD scale evolution of quantum entanglement in double quark states in
the proton. These results are qualitative because
we do not resum multiple evolution steps and are therefore limited
to unrealistically weak coupling, $\alpha_s=0.1$, and large cutoff
on the LC momentum of the gluon, $x=0.1$. Even so, interestingly, at the higher scale we do observe
significant effects due to quantum correlations for nearly symmetric
momentum fractions, $x_1 \simeq x_2$, for fairly small $x_1+x_2
\lesssim 0.3$, where their effect is to suppress the dPDF. These
initial explorations motivate the development of the complete QCD
evolution equations of the full two-quark density matrix, to account
for the emission of an entire ladder of partons. This work is in
progress.

\section{Acknowledgements}

We acknowledge support by the DOE Office of Nuclear Physics through Grant DE-SC0002307.

\bibliography{refs}

\end{document}